\documentclass[12pt]{article}
\input{epsf.sty}

\def\fo{\hbox{{1}\kern-.25em\hbox{l}}}

\def\slashchar#1{\setbox0=\hbox{$#1$}           
   \dimen0=\wd0                                 
   \setbox1=\hbox{/} \dimen1=\wd1               
   \ifdim\dimen0>\dimen1                        
      \rlap{\hbox to \dimen0{\hfil/\hfil}}      
      #1                                        
   \else                                        
      \rlap{\hbox to \dimen1{\hfil$#1$\hfil}}   
      /                                         
   \fi}                                         %

\def\hide#1{[hidden stuff]}

\def\beq{\begin{equation}}
\def\eeq{\end{equation}}
\def\eq{\end{equation}}
\def\to{\rightarrow}

\def\mZ{m_Z}

\def\cbeta{c_{\beta}}

\def\stilde{\widetilde}
\def\sbeta{s_{\beta}}
\def\cW{c_{W}}   
\def\sW{s_{W}}   
\def\mEt{\mbox{${\hbox{$E$\kern-0.6em\lower-.1ex\hbox{/}}}_T$}\, } 

\def\bsg{\ifmmode B\to X_s\gamma\else $B\to X_s\gamma$\fi}
\def\bsll{\ifmmode B\to X_s\ell^+\ell^-\else $B\to X_s\ell^+\ell^-$\fi}
\def\bstt{\ifmmode B\to X_s\tau^+\tau^-\else $B\to X_s\tau^+\tau^-$\fi}
\def\shat{\ifmmode \hat{s}\else $\hat{s}$\fi}

\newcommand{\newc}{\newcommand}

\newc{\asusy}{\delta a^{\rm SUSY}_\mu}
\newc{\lcal}{\int {\cal L}dt}
 
\newc{\LSP}{{\chi^0_1}}
\newc{\stauR}{{\tilde \tau_R}}
\newc{\stau}{{\tilde \tau_1}}
\newc{\mstop}{m_{\tilde{t}}}
\newc{\mHpm}{m_{H^\pm}}
\newc{\gsim}{\lower.7ex\hbox{$\;\stackrel{\textstyle>}{\sim}\;$}}
\newc{\lsim}{\lower.7ex\hbox{$\;\stackrel{\textstyle<}{\sim}\;$}}
\newc{\ie}{{\it i.e.}}          
\newc{\etal}{{\it et al.}}
\newc{\eg}{{\it e.g.}}          
\newc{\kev}{\hbox{\rm\,keV}}            
\newc{\mev}{\hbox{\rm\,MeV}}            
\newc{\gev}{\hbox{\rm\,GeV}}            
\newc{\tev}{\hbox{\rm\,TeV}}
\newc{\xpb}{\hbox{\rm\, pb}}
\newc{\xfb}{\hbox{\rm\, fb}}

\newc{\mtop}{m_t}
\newc{\mbot}{m_b}
\newc{\mz}{m_Z}
\newc{\mw}{M_W}
\newc{\alphasmz}{\alpha_s(m_Z^2)}
\newc{\swsq}{\sin^2\theta_W}
\newc{\tw}{\tan\theta_W}
\newc{\cw}{\cos\theta_W}
\newc{\sw}{\sin\theta_W}
\newc{\BR}{\hbox{\rm BR}}
\newc{\zbb}{Z\to b\bar}
\newc{\Gb}{\Gamma (Z\to b\bar b)}
\newc{\Gh}{\Gamma (Z\to \hbox{\rm hadrons})}
\newc{\rbsm}{R_b^\hbox{\rm sm}}
\newc{\rbsusy}{R_b^\hbox{\rm susy}}
\newc{\drb}{\delta R_b}

\newc{\sgn}{\mbox{sgn}}
\newc{\tbeta}{\tan\beta}
\newc{\uL}{{\tilde u_L}}
\newc{\uR}{{\tilde u_R}}
\newc{\cL}{{\tilde c_L}}
\newc{\cR}{{\tilde c_R}}
\newc{\tL}{{\tilde t_L}}
\newc{\tR}{{\tilde t_R}}
\newc{\dL}{{\tilde d_L}}
\newc{\dR}{{\tilde d_R}}
\newc{\sL}{{\tilde s_L}}
\newc{\sR}{{\tilde s_R}}
\newc{\bL}{{\tilde b_L}}
\newc{\bR}{{\tilde b_R}}
\newc{\eL}{{\tilde e_L}}
\newc{\eR}{{\tilde e_R}}
\newc{\mhp}{m_{H^\pm}}
\newc{\mhalf}{m_{1/2}}
\newc{\emt}{{e/\mu /\tau}}

\newc{\lR}{\tilde{l}_R}
\newc{\lL}{\tilde{l}_L}
\newc{\nL}{\tilde{\nu}_L}
\newc{\na}{\chi^0_1}
\newc{\nb}{\chi^0_2}
\newc{\nc}{\chi^0_3}
\newc{\nd}{\chi^0_4}
\newc{\ca}{\chi^{\pm}_1}
\newc{\cb}{\chi^{\pm}_2}
\newc{\camp}{\chi^\mp_1}
\newc{\cbmp}{\chi^\mp_1}
\newc{\capos}{\chi^{+}_1}
\newc{\caneg}{\chi^{-}_1}
\newc{\phit}{\phi_t}
\newc{\phib}{\varphi_b}
\newc{\phiew}{\phi_{ew}}
\newc{\htz}{h^0_t}
\newc{\hbz}{h^0_b}
\newc{\hewz}{h^0_{ew}}
\newc{\hsmz}{h^0_{sm}}
\newc{\huz}{h^0_u}
\newc{\hsusyz}{h^0_{susy}}

\newc{\C}{{\cal C}}

\newcommand{\drawsquare}[2]{\hbox{%
\rule{#2pt}{#1pt}\hskip-#2pt
\rule{#1pt}{#2pt}\hskip-#1pt
\rule[#1pt]{#1pt}{#2pt}}\rule[#1pt]{#2pt}{#2pt}\hskip-#2pt
\rule{#2pt}{#1pt}}

\newc{\Dal}{\drawsquare{7}{0.6}}

\def\dofig#1#2{\epsfxsize=#1\centerline{\epsfbox{#2}}}
\def\dofigs#1#2#3{\centerline{\epsfxsize=#1\epsfbox{#2}%
   \hfil\epsfxsize=#1\epsfbox{#3}}}

\def\beq{\begin{equation}}
\def\eeq{\end{equation}}
\def\bea{\begin{eqnarray}}
\def\eea{\end{eqnarray}}
\catcode`@=11
\long\def\@caption#1[#2]#3{\par\addcontentsline{\csname
  ext@#1\endcsname}{#1}{\protect\numberline{\csname
  the#1\endcsname}{\ignorespaces #2}}\begingroup
    \small
    \@parboxrestore
    \@makecaption{\csname fnum@#1\endcsname}{\ignorespaces #3}\par
  \endgroup}
\catcode`@=12

\begin{document}
\begin{titlepage}

\begin{flushright}
hep-ph/0103067 \\
Fermilab-Pub-01/030-T \\
LBNL-47586 \\
\end{flushright}

\huge
\bigskip
\bigskip
\begin{center}
{\Large\bf
Muon anomalous magnetic dipole moment}\\
{\Large\bf in supersymmetric theories}
\end{center}

\large

\vspace{.15in}
\begin{center}

Stephen P.~Martin$^a$ and James D.~Wells$^b$

\small

\vspace{.1in}
{\it $^{(a)}$Department of Physics, Northern Illinois University,
         DeKalb IL 60115 {\rm and} \\
Fermi National Accelerator Laboratory, PO Box 5000, Batavia IL 60510 \\}
\vspace{0.1cm}
{\it $^{(b)}$Physics Department, University of California, 
      Davis CA 95616, USA {\rm and} \\
 Theory Group, Lawrence Berkeley National Laboratory, Berkeley CA 94720}

\end{center}
 
\vspace{0.15in}
 
\begin{abstract}

We study the muon anomalous magnetic dipole moment in supersymmetric theories.
The impact of the recent Brookhaven E821 experimental measurement on both
model-independent and model-dependent supersymmetric parameter spaces is
discussed in detail. We find that values of $\tan\beta$ as low as 3 can be
obtained while remaining within the E821 one-sigma bound. This requires a
light smuon; however, we show that, somewhat surprisingly, no
model-independent bound can be placed on the mass of the lightest chargino
for any $\tan\beta \geq 3$.  We also show that the maximum contributions
to the anomalous magnetic moment are insensitive to CP-violating phases.
We provide analyses of the supersymmetric contribution to the muon
anomalous magnetic moment in dilaton-dominated supergravity models and
gauge-mediated supersymmetry-breaking models. Finally, we discuss how
other phenomena, such as $B(b\to s\gamma)$, relic abundance of the
lightest superpartner, and the Higgs mass may be correlated with the
anomalous magnetic moment, but do not significantly impact the viability
of a supersymmetric explanation, or the mass limits obtainable on smuons
and charginos.

\end{abstract}

\medskip


\end{titlepage}

\baselineskip=18pt
\setcounter{footnote}{1}
\setcounter{page}{2}
\setcounter{figure}{0}
\setcounter{table}{0}
\tableofcontents

\section{The muon anomalous magnetic dipole moment}
\setcounter{equation}{0}
\setcounter{footnote}{1}

\subsection{Standard Model prediction and experiment}

The amplitude for the photon-muon-muon coupling in the limit of
the photon momentum $q$ going to zero can be written as
\bea
{\rm Amplitude}=ie\, \bar u \left[ \gamma^\lambda + 
    a_\mu\frac{i\sigma^{\lambda\beta} q_\beta}{2m_\mu} \right] u\, A_\lambda,
\eea
where $e=\sqrt{4\pi\alpha_{\rm EM}}$.  The second term comes from
loop corrections, and is given to
one-loop order in QED by $a_\mu = \frac{\alpha}{2\pi}$.   Being 
a small correction to the tree-level magnetic moment of the muon, it is
called the anomalous magnetic moment.  

The state of the art calculation of $a_\mu$ within the Standard Model (SM)
is~\cite{Czarnecki:2001pv}
\bea
a^{\rm SM}_\mu = 11\, 659\, 159.6(6.7)\times 10^{-10}.
\eea
The majority of the uncertainty comes from hadrons in the photon vacuum
polarization diagram.

Recently the Brookhaven E821 experiment has released a new measurement
of $a_\mu$ and found~\cite{Brown:2001mg}
\bea
a^{\rm E821}_\mu = 11\, 659\, 202(14)(6)\times 10^{-10}.
\eea
From this one concludes~\cite{Brown:2001mg}
\bea
\delta a_\mu =a^{\rm E821}_\mu - a^{\rm SM}_\mu = (43 \pm 16)\times
10^{-10}.
\eea
This result indicates that the anomalous magnetic moment
of the muon may need additional contributions beyond the SM to
be consistent with the experimental measurement.  

\subsection{Supersymmetric contributions}

There are many reasons to believe that the SM is an
incomplete description of nature besides the present indications
from $a_\mu$.  For example, the SM does not explain 
baryogenesis, dark matter, the ratios of fundamental scales, or the
strengths of gauge and Yukawa interactions.  Supersymmetry is an 
appealing theoretical framework that may answer many of the
questions unanswerable within the SM~\cite{Haber:1985rc,Martin:1997ns}.  

The supersymmetry 
effects~\cite{fayet}%
-\cite{Brignole:1999gf} 
on $a_\mu$ 
include loops with a chargino and a
muon sneutrino, and loops with a neutralino and a smuon.  Summations
are performed over all such chargino, neutralino and smuon mass eigenstates.
The one-loop superpartner contributions to $a_\mu$, including the effects
of possible complex phases, are
\bea
\delta a_\mu^{\chi^0} & = & \frac{m_\mu}{16\pi^2}
   \sum_{i,m}\left\{ -\frac{m_\mu}{ 12 m^2_{\tilde\mu_m}}
  (|n^L_{im}|^2+ |n^R_{im}|^2)F^N_1(x_{im}) 
 +\frac{m_{\chi^0_i}}{3 m^2_{\tilde \mu_m}}
    {\rm Re}[n^L_{im}n^R_{im}] F^N_2(x_{im})\right\}\phantom{xxxx}\\
\delta a_{\mu}^{\chi^\pm} & = & \frac{m_\mu}{16\pi^2}\sum_k
  \left\{ \frac{m_\mu}{ 12 m^2_{\tilde\nu_\mu}}
   (|c^L_k|^2+ |c^R_k|^2)F^C_1(x_k)
 +\frac{2m_{\chi^\pm_k}}{3m^2_{\tilde\nu_\mu}}
         {\rm Re}[ c^L_kc^R_k] F^C_2(x_k)\right\}\phantom{xxxx}
\eea
where $i=1,2,3,4$ and $m=1,2$ and $k=1,2$ are neutralino and smuon
and chargino mass eigenstate labels respectively, and
\bea
n^R_{im} & = &  \sqrt{2} g_1 N_{i1} X_{m2} + y_\mu N_{i3} X_{m1}\\
n^L_{im} & = &  {1\over \sqrt{2}} \left (g_2 N_{i2} + g_1 N_{i1}
\right ) X_{m1}^* - y_\mu N_{i3} X^*_{m2}\\
c^R_k & = & y_\mu U_{k2} \\
c^L_k & = & -g_2V_{k1}
\eea
and $y_\mu = g_2 m_\mu/\sqrt{2} m_W \cos\beta$ is the muon Yukawa
coupling.
The kinematic loop functions depend on the
variables $x_{im}=m^2_{\chi^0_i}/m^2_{\tilde\mu_m}$, 
$x_k=m^2_{\chi^\pm_k}/m^2_{\tilde\nu_\mu}$ and are given by
\bea
F^N_1(x) & = &\frac{2}{(1-x)^4}\left[ 1-6x+3x^2+2x^3-6x^2\ln x\right] \\
F^N_2(x) & = &\frac{3}{(1-x)^3}\left[ 1-x^2+2x\ln x\right] \\
F^C_1(x) & = &\frac{2}{(1-x)^4}\left[ 2+ 3x - 6
x^2 + x^3
        +6x\ln x\right] \\
F^C_2(x) & = & -\frac{3}{2(1-x)^3}\left[ 3-4x+x^2
    +2\ln x\right],
\eea
normalized so that $F^N_1(1) = F^N_2(1) = F^C_1(1) = F^C_2(1) = 1$,
corresponding to degenerate sparticles.

By definition
$g_2\simeq 0.66$ and $g_1\simeq 0.36$ are the $SU(2)_L$ and $U(1)_Y$
gauge couplings. The phase convention for $\mu$ follows 
Refs.~\cite{Haber:1985rc,Martin:1997ns}, so that the neutralino
and chargino mass matrices are given by
\beq
M_{\chi^0} = \pmatrix{M_1 & 0 & - \cbeta\, \sW\, \mZ &
\sbeta\, \sW \, \mZ\cr
0 & M_2 & \cbeta\, \cW\, \mZ & - \sbeta\, \cW\, \mZ \cr
-\cbeta \,\sW\, \mZ & \cbeta\, \cW\, \mZ & 0 & -\mu \cr
\sbeta\, \sW\, \mZ & - \sbeta\, \cW \, \mZ& -\mu & 0 \cr }
\label{neutralinomassmatrix} 
\eeq
and
\beq
M_{\chi^\pm} = 
\pmatrix{M_2 & \sqrt{2} \sbeta\, m_W\cr
                              \sqrt{2} \cbeta\, m_W & \mu \cr }.
\label{charginomassmatrix}
\eeq
Here we have used abbreviations $\sbeta = \sin\beta$,
$\cbeta = \cos\beta$, $\sW = \sin\theta_W$, and $\cW = \cos\theta_W$.
The
neutralino
mixing matrix 
$N_{ij}$ 
and the chargino mixing  matrices $U_{kl}$ and $V_{kl}$
are identical to those in
Refs.~\cite{Haber:1985rc,Martin:1997ns}; they satisfy
\bea
N^* M_{\chi^0} N^\dagger &=& {\rm diag}(
m_{\chi^0_1},
m_{\chi^0_2},
m_{\chi^0_3},
m_{\chi^0_4})\\
U^* {M}_{\chi^\pm} V^\dagger &=& {\rm diag}(
m_{\chi^\pm_1},
m_{\chi^\pm_2})
. 
\eea
In
particular, the neutralino and chargino mass eigenvalues are always
chosen to be real and positive, regardless of the complex phases of the
underlying Lagrangian parameters; all non-trivial phases are contained
in the unitary mixing matrices $N$, $U$, $V$. 
The smuon mass matrix, written in the 
$\{ \tilde\mu_L, \tilde\mu_R \}$ basis is
\bea
M^2_{\tilde\mu}=\left( \begin{array}{cc}
   m^2_L +(s_W^2 -\frac{1}{2})m_Z^2\cos\,  2\beta &
              m_\mu (A^*_{\tilde\mu}-\mu\tan\beta) \\
 m_\mu (A_{\tilde\mu}-\mu^*\tan\beta) &
    m^2_R -s_W^2 \, m_Z^2\cos\,  2\beta 
\end{array}\right),
\eea
and the unitary matrix $X_{mn}$ is defined by
\bea
X M^2_{\tilde\mu}\, X^\dagger = 
{\rm diag}\, (m^2_{\tilde\mu_1}, m^2_{\tilde\mu_2}).
\eea
The muon sneutrino mass is related to the left-handed smuon mass parameter
by
\beq
m_{\stilde \nu}^2 = m_L^2 + {1\over 2} m_Z^2 \cos 2\beta  .
\eeq

The simplest analytic result to obtain from supersymmetry is to assume that
all superpartners have the same mass $M_{\rm SUSY}$, which leads to
\bea
\delta a_\mu^{\rm SUSY} =\frac{\tan\beta}{192\pi^2}
    \frac{m_\mu^2}{M_{\rm SUSY}^2}\, (5g_2^2+g_1^2)
=
14 \tan\beta \left ( \frac{100\>{\rm GeV}}{M_{\rm SUSY}} \right )^2
10^{-10}
\label{msusy}
\eea
with the chargino contribution dominating the neutralino
contribution~\cite{Moroi:1996yh}.  The large $\tan\beta$ scaling
is easy to understand, and is analogous to the large
$\tan\beta$ enhancements of $B(b\to s\gamma)$ and $\Delta m_b$ corrections.
$a_\mu$ requires a muon chirality flip, which usually costs a  $m_\mu$
suppression.  However, the higgsino-smuon-muon 
vertex coupling can perform the
chirality flip with the muon Yukawa coupling $y_\mu$, leading to
an enhancement $y_\mu\propto m_\mu\tan\beta$ at large $\tan\beta$.

Another important limit which will play a role in the discussion of
the next section is the case in which $M_1 \ll M_2, \mu$, so that
only loops containing a light bino and the smuons are important.
In that limit, we find:
\bea
\delta a_\mu^{{\rm light}\> {\rm bino}} &=&
{g_1^2 \over 48 \pi^2} {m^2_\mu  M_1 {\rm Re} [\mu \tan\beta - A_\mu^*]
\over m_{\stilde \mu_2}^2 - m_{\stilde \mu_1}^2 }
\left[ {F_2^N(x_{11}) \over m_{\stilde \mu_1}^2} 
- {F_2^N(x_{12})\over m^2_{\stilde \mu_2}} \right ]
\label{lightbino}
\eea
where $x_{1m} = M_1^2/m_{\stilde \mu_m}^2$. Note that eq.~(\ref{lightbino})
has a smooth limit as the sleptons become degenerate.
This yields a quite sizeable contribution in the case that
all neutralinos and charginos except the light bino become heavy.
For example, in the case $m_{\stilde \mu_1} \approx m_{\stilde \mu_2} =
2.0 M_1$, eq.~(\ref{lightbino}) becomes
\bea
\delta a_\mu^{{\rm light}\> {\rm bino}} & = &
18 \tan\beta \left ( {100\>{\rm GeV} \over m_{\stilde \mu} } \right )^3
\left ( {\mu - A_\mu \cot\beta \over 1000 \>{\rm GeV}} \right ) 10^{-10}
\label{nouseforanumber}
\eea
(This formula will eventually fail to be accurate
for extremely huge $\mu \tan\beta$, in accord with decoupling,
since then $m_{\stilde \mu_1} \approx m_{\stilde
\mu_2}$ must fail badly.)
This situation is not quite in effect in the usual supergravity-inspired
and gauge-mediated supersymmetry breaking scenarios, but is certainly
obtainable within a model-independent framework, as we shall see.
Furthermore, it could arise quite naturally in certain well-defined
extensions of the MSSM. For example, if supersymmetry breaking is
manifested by an $F$-term VEV transforming in the adjoint 
{\bf 24} representation of a GUT $SU(5)$ gauge group, then the  
gaugino mass parameters are in the approximate ratio $M_1$ : $M_2$ : $M_3$ ::
$1$ : $6$ : $-12$ at the electroweak scale \cite{Anderson:2000ui,Abel:2000vs}. 
Another class of examples occurs in gauge-mediated supersymmetry breaking
\cite{Giudice:1999bp,Culbertson:2000am}
(GMSB) models which have messengers that are not in complete
$SU(5)$ multiplets, but rather in representations with more
electroweak singlets than doublets. These models naturally
predict a bino and right-handed
sleptons which are much lighter than all electroweak doublet superpartners
\cite{Martin:1997zb}.

The leading log contribution from two-loop evaluation~\cite{Degrassi:1998es} 
yields a suppression
\beq
a^{\rm SUSY}_{\mu,\, 2\,{\rm loop}}=a^{\rm SUSY}_{\mu,\, 1\,{\rm loop}}\,
\left( 1- \frac{4\alpha}{\pi}\ln
\frac{M_{\rm SUSY}}{m_\mu} \right)
\eeq
where $M_{\rm SUSY}$ is a typical superpartner mass.
This suppression factor
varies between about 7\% and 9\% for the parameter space we consider.
Although a complete NLO calculation has yet to be carried through
in supersymmetry, 
we have imposed in all of our numerical results below 
a uniform 7\% reduction from the 1-loop calculation
based on this leading-log estimate.

\section{Results for general supersymmetric models}
\setcounter{equation}{0}
\setcounter{footnote}{1}

\subsection{General MSSM parameters}
The 
 full Minimal Supersymmetric Standard Model (MSSM) 
parameter space contains dozens of parameters. However,
the supersymmetric contribution to the muon anomalous magnetic moment
depends at tree-level only on the quantities $M_1$, $M_2$, $\mu$,
$\tan\beta$, $m_{L}^2$, $m_{R}^2$, and $A_\mu$. 
Therefore
it is possible to comprehend the impact of supersymmetry by
using scans over parameter space which include experimental
constraints. Several recent papers have examined
the question of whether bounds can be put on superpartner masses and
other parameters by taking the E821 results at face value. In this
section, we remark on the possibility of extracting such bounds
in a model-independent, and therefore maximally conservative,
supersymmetric framework.

We have conducted an exhaustive examination of the relevant MSSM parameter
space without imposing conditions that follow from model-building
prejudice, in particular 
without requiring the usual gaugino mass unification condition
between $M_1$ and $M_2$. In general, the supersymmetric
contribution to $a_\mu$ can be made larger for larger values of
$\tan\beta$ and smaller masses of the lighter chargino. However,
in contrast to some recent reports, we find
that it is quite possible to accommodate the E821 results even with
rather low $\tan\beta$ and for arbitrarily heavy charginos. This can
be seen directly from eq.~(\ref{lightbino}), by plugging in typical
values.
As long as $M_1$ and $M_2$ are
not tied together by a unification condition, the charginos can
become very heavy 
for very large $M_2$ and $\mu$ while still leaving behind a
contribution which is large enough to fall within the E821
1-$\sigma$ bounds, provided only that a smuon is light and $|\mu |$
is not too small.
Even for $\tan\beta=2$, the contribution can be large enough to fall within
the present 2-$\sigma$ bounds.

\begin{figure}[tpb]
\dofigs{3.5in}{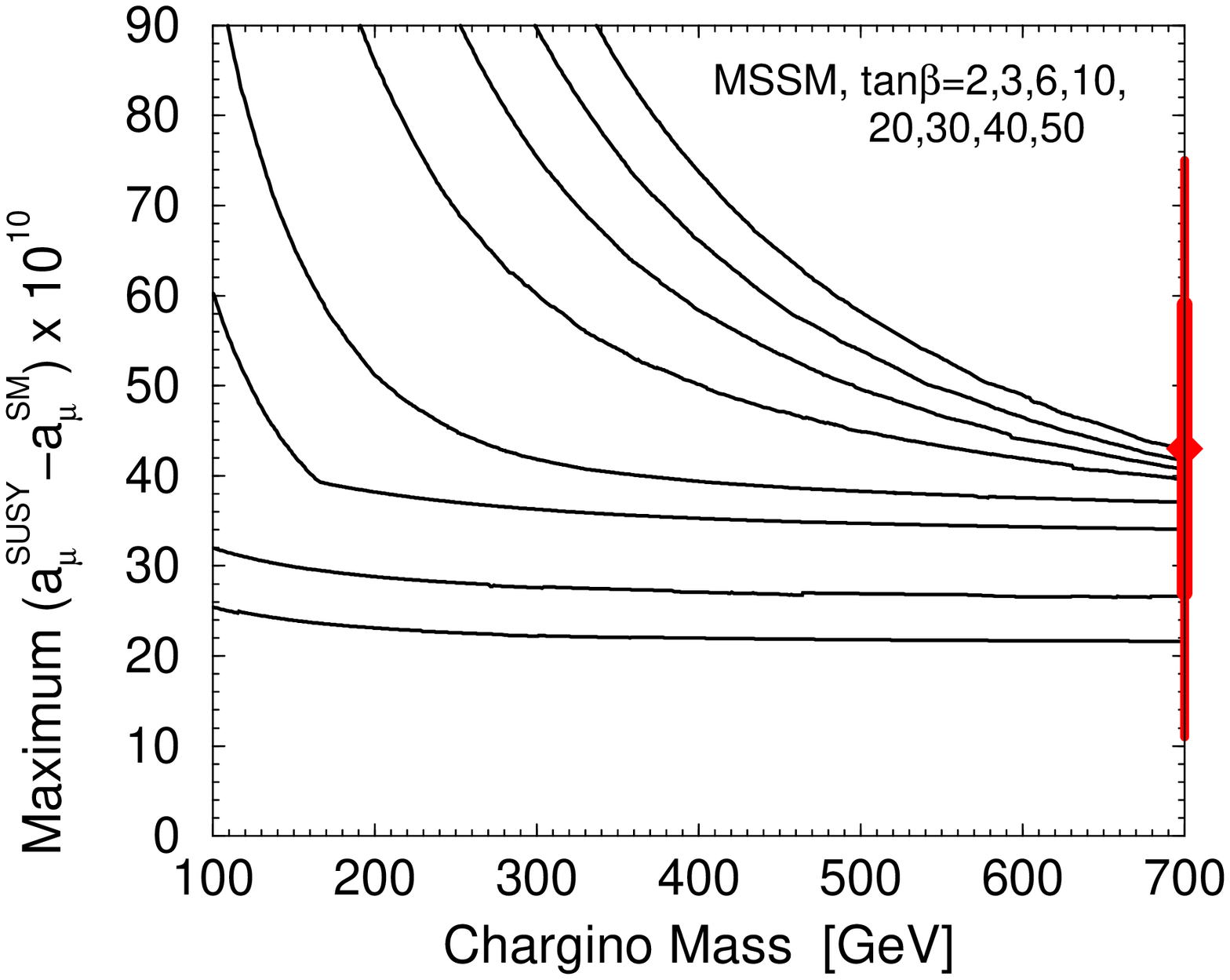}{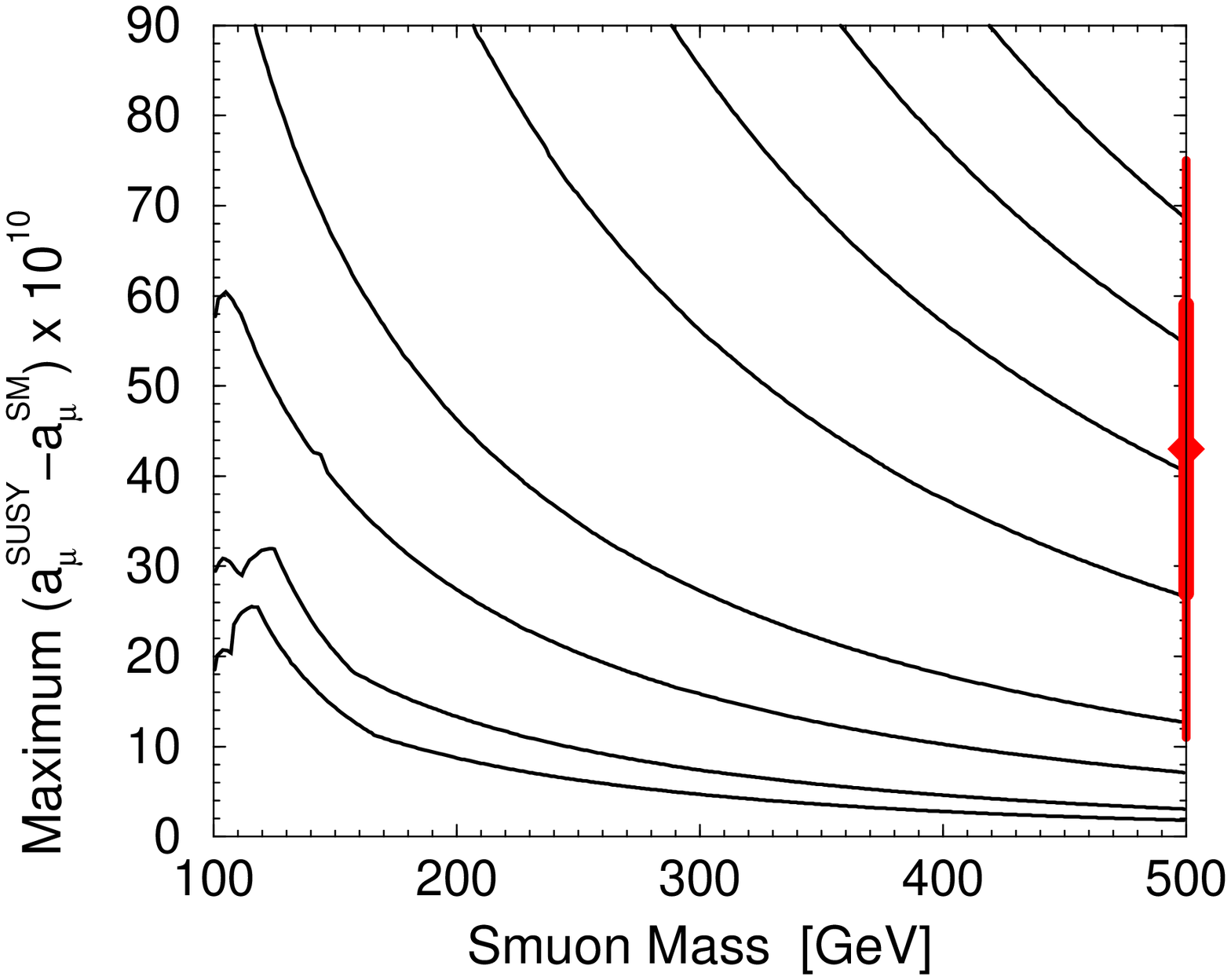}
\caption{The maximum possible values for the
supersymmetric contribution to the muon anomalous magnetic 
moment, as a function of the lighter chargino mass and the 
lighter smuon mass. Gaugino mass unification conditions
have not been imposed. All charged superpartners
are required to be heavier than 100 GeV, and the lightest
neutralino is required to be heavier than 50 GeV. The maximum
allowed value for $|\mu|$ is taken to be 1000 GeV. 
The contours are shown from bottom to top for $\tan\beta =$
2, 3, 6, 10, 20, 30, 40, 50. The red bars on the right vertical
axes indicate the 1-$\sigma$ and 2-$\sigma$ allowed regions from
the Brookhaven E821 experiment.
\label{maxgeneral}}
\end{figure}

The results for the maximum possible value of 
$a_\mu^{\rm SUSY} - a_\mu^{\rm SM}$ 
are shown in fig.~\ref{maxgeneral}.
Here we have chosen to present contours for different values of
$\tan\beta = $ 2, 3, 6, 10, 20, 30, 40, 50. All of the other
parameters are
taken to be independent, subject to the constraint that all charged
superpartners are heavier than 100 GeV and the lightest neutralino
is heavier than  50 GeV (to
provide approximate agreement
with present and imminent bounds from the final results of the LEP2
experiments). In order to avoid the possibility of charge- and
color-breaking vacua, we have imposed a constraint 
$|A_\mu| \leq 3 {\rm Min}[m_{L}, m_{R}]$. 
The precise value of this
bound generally does not
have a large effect on the contours shown. 
However, in these plots we have found it appropriate to make 
two significant concessions to model-dependent prejudice, as follows.
 
First, we have also included a 100 GeV lower bound 
on the lighter stau mass, by
assuming universality in soft slepton parameters  
$A_{\stilde \tau} = A_{\stilde \mu}$ and  
$m_{L,\stilde \tau}^2 =  m_{L,\stilde \mu}^2$ and 
$m_{R,\stilde \tau}^2 =  m_{R,\stilde \mu}^2$, 
and then requiring $m_{\stilde \tau_1} > 100$ GeV.  With these assumptions,
requiring the staus to be heavier than 100 GeV imposes a stronger indirect
constraint on the smuon, because of the mixings proportional to
$m_\tau \mu \tan\beta$ for staus and $m_{\mu} \mu \tan\beta$ for smuons.
Strictly
speaking, this type of requirement does not correspond to a
model-independent framework, where lepton flavor universality need not be
imposed; all bounds from low-energy lepton-number violation can be evaded
by simple alignment in lepton flavor space. However, perhaps the most
natural
way to satisfy these constraints is to impose lepton universality at high
energies. This constraint is most significant for smaller smuon masses
and smaller values of
$\tan\beta$ (less than roughly 10 or so), where the chargino-sneutrino 
loops do not necessarily dominate in $a_\mu^{\rm SUSY} - a_\mu^{\rm SM}$.  
Since
in most cases the bounds are saturated by large mixing in the slepton
sectors arising dominantly from the effects of large $\mu$,  
this requirement is not
very sensitive to the precise values used for the soft
parameters $A_{\stilde \tau}$, $m_{R,\stilde\tau}$ and
$m_{L,\stilde \tau}$, which can be affected by renormalization
group running from a high scale where universality is imposed.  
In any case, we emphasize
that in principle {\it even larger} 
values of  $a_\mu^{\rm SUSY} - a_\mu^{\rm SM}$ 
can be obtained than are presented here.

Second, we have imposed a maximum value of $|\mu| < 1000$ GeV. 
If one chooses to allow larger values of $\mu$, then
one can construct models with larger contributions to
$a_\mu^{\rm SUSY}$, resulting from neutralino-smuon loops
dominated by a light smuon with a large mixing angle
due to the off-diagonal terms proportional to $\mu$ in the squared-mass 
matrix. The prospect of very large
$|\mu |$ often causes discomfort since it
requires fine-tuning in the Higgs potential in order to obtain
electroweak symmetry breaking in accord with experiment.
However, it should be noted that in general the upper bound on contributions
to $a_\mu^{\rm SUSY}$ increases with the assumed maximum allowed $|\mu |$.

We have not imposed any requirement that the lightest supersymmetric
particle (LSP) is a neutralino. While the existence of 
a neutralino LSP could make an
attractive candidate for the cold dark matter, in a general model
framework it is
neither a necessary
nor a sufficient condition for an acceptable cosmology. 
Furthermore, in models which saturate
the maximum possible $a^{\rm SUSY}_\mu$, a neutralino is typically 
light, so that  imposing such a constraint would
generally not affect our results, except below when we impose gaugino
mass unification on the parameter space.

Several features of fig.\ \ref{maxgeneral} deserve comment. In
the graph of maximum $a_\mu^{\rm SUSY} - a_\mu^{\rm SM}$ as a function of
chargino mass, there are distinct regions separated by an ``elbow"
(which is most visibly pronounced in the case of $\tan\beta=6$).
For chargino masses to the right of 
the elbow in each case, the bound is saturated
by models with the maximum allowed value of $|\mu |$ and small $|M_1|$, 
and in fact the graph
is nearly flat as the dominant contribution comes mainly from
neutralino-smuon loops, as in eq.~(\ref{lightbino}). 
These models also have smuon masses (and stau
masses) near their lower bound. For chargino masses to the left of 
the elbow,
the maximum of $a_\mu^{\rm SUSY} - a_\mu^{\rm SM}$ tends to be saturated for
models with much smaller values of $|\mu|$, and the chargino loops
play a more important role. As $\tan\beta$ is increased, the chargino
loops become relatively more important, and the dependence on the chargino
masses extends out to much larger values before they decouple.

In the graph of maximum $a_\mu^{\rm SUSY} - a_\mu^{\rm SM}$ as a function of
smuon mass in fig.\ \ref{maxgeneral}, the cases with smaller $\tan\beta$
exhibit some structure. For smuon masses just above 100 GeV, the models
that saturate the bound have $|\mu|$ of order 200 GeV; much 
larger values of
$|\mu |$ which could otherwise increase  $a_\mu^{\rm SUSY}$ 
would conflict with our assumptions stated above regarding the
limit on the lighter stau mass. In an intermediate region for the lightest
smuon mass, the models that saturate the bound are the ones with
the maximum allowed value of $|\mu|$ and small $|M_1|$, 
as suggested by eq.~(\ref{lightbino}). 
This leads to a bump in the maximum
$a_\mu^{\rm SUSY} - a_\mu^{\rm SM}$; this is prominent for
$\tan\beta=2,3$, is just barely visible for $\tan\beta=6$ (near smuon
mass of 145 GeV)
and disappears entirely for larger values of $\tan\beta$. 
For larger $\tan\beta$ or larger smuon masses, the models that saturate
the bound again have much smaller $|\mu|$ (of order 200 GeV).

\begin{figure}[tpb]
\dofigs{3.5in}{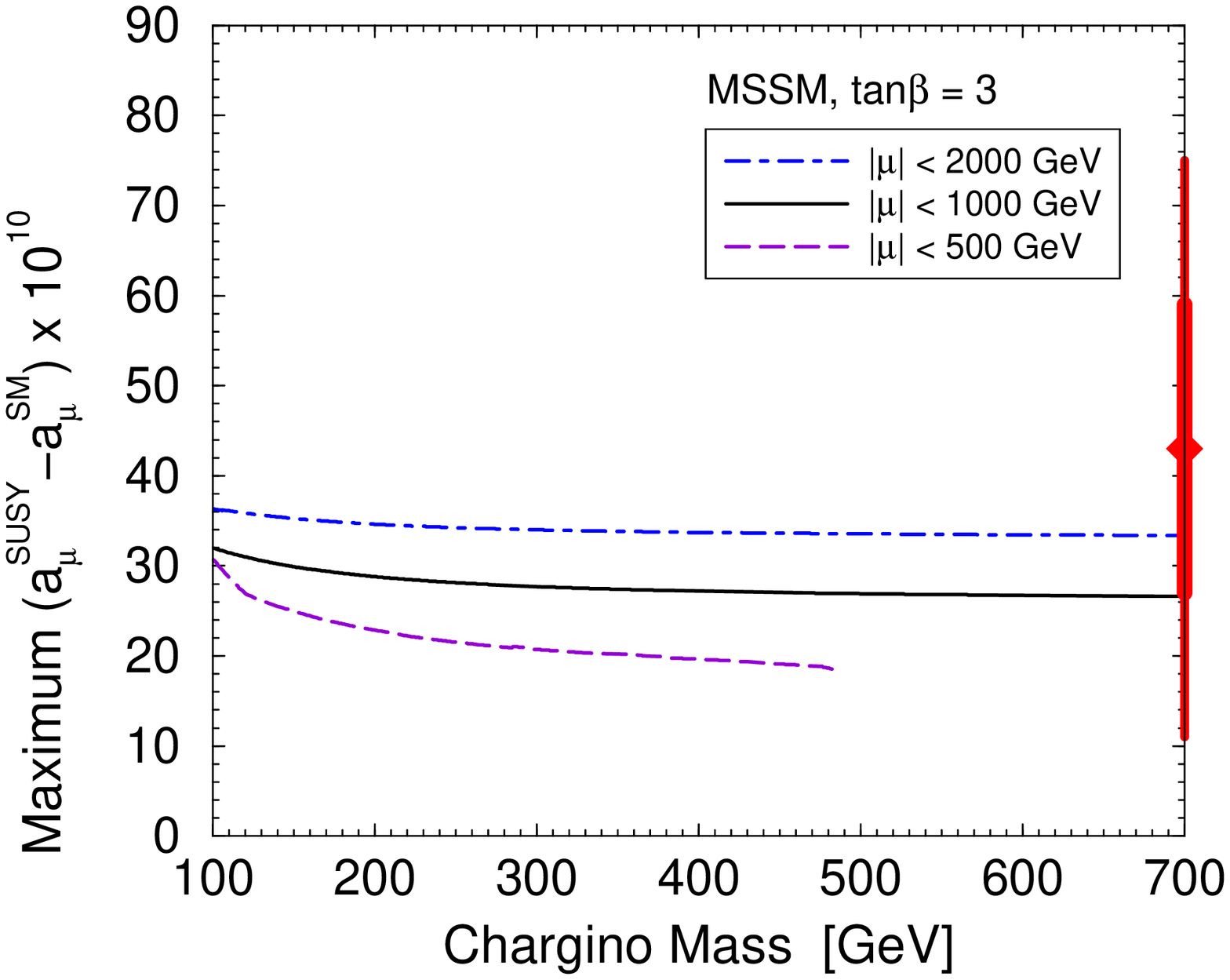}{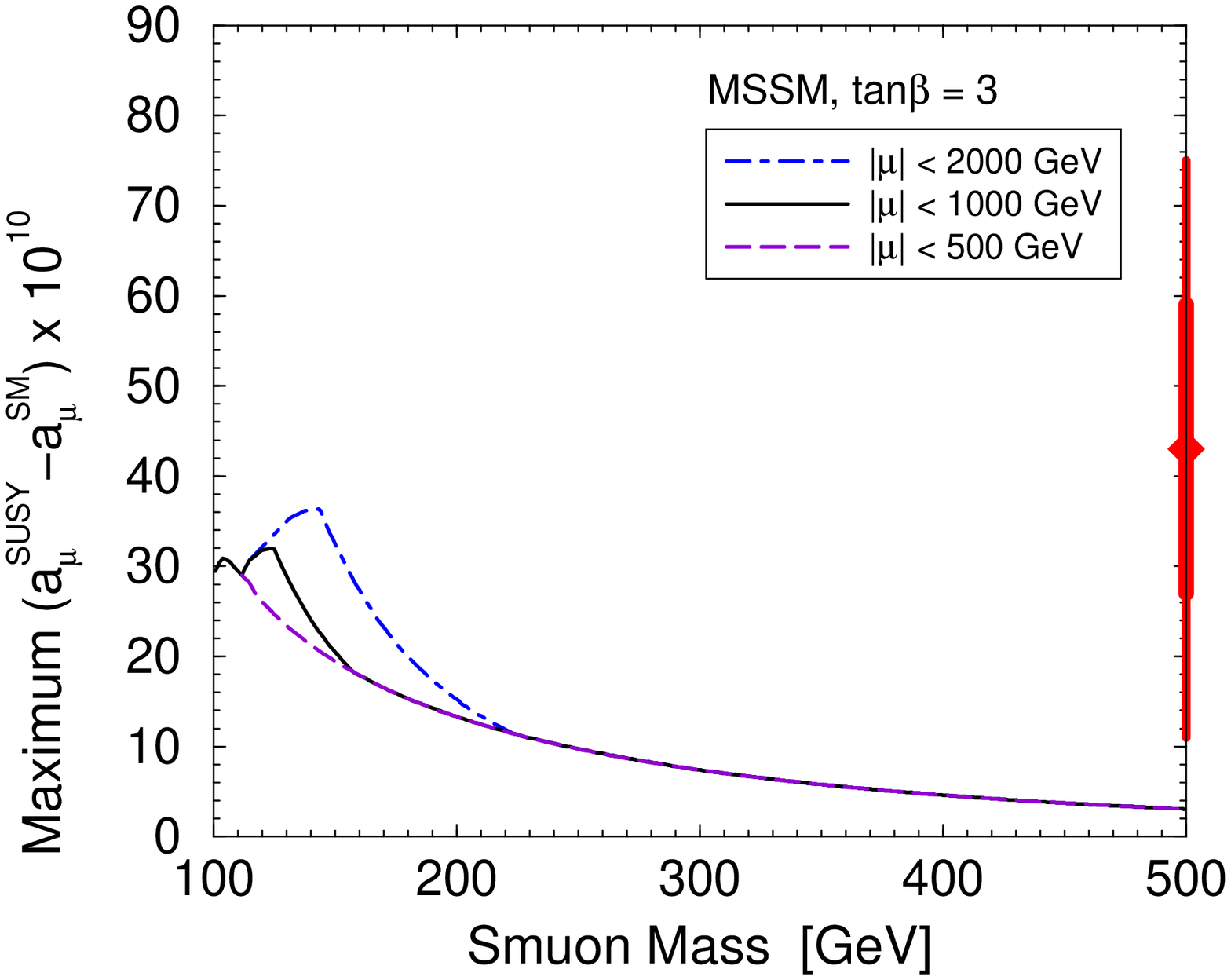}
\caption{Effects of larger allowed $|\mu |$: 
the maximum possible values for the
supersymmetric contribution to the muon anomalous magnetic 
moment, as a function of the lighter chargino mass and the lighter
smuon mass, for $\tan\beta=3$ and
different values (2000 GeV, 1000 GeV, 500 GeV from top to bottom)
of the allowed maximum $|\mu|$.  Gaugino mass unification conditions
have not been imposed. All charged superpartners
are required to be heavier than 100 GeV, and the lightest
neutralino is required to be heavier than 50 GeV.
The red bars on the right vertical
axes indicate the 1-$\sigma$ and 2-$\sigma$ allowed regions from
the Brookhaven E821 experiment.
\label{maxgeneralmuvar}}
\end{figure}

The effect of varying the maximum allowed value for $|\mu|$ is 
illustrated in fig.~\ref{maxgeneralmuvar} for
$\tan\beta=3$, using $|\mu| < 500, 1000,$ and 2000 GeV.
The graph shows that for a given chargino mass, the upper bound on
$a_\mu^{\rm SUSY} - a_\mu^{\rm SM}$ is usually 
obtained for the maximum allowed $|\mu|$.
However, as a function of the lighter smuon mass, the upper bound on
$a_\mu^{\rm SUSY} - a_\mu^{\rm SM}$ is saturated for large $|\mu|$ 
only in a finite range of the smuon mass. Again, 
this is because for smuon masses
very close to the experimental limit, the effects of large $|\mu|$ are
limited by our requirement that the stau is not too light, while for
sufficiently large smuon masses the chargino-sneutrino loops become
more important.

\subsection{Gaugino mass unification}

\begin{figure}[tpb]
\dofig{3.75in}{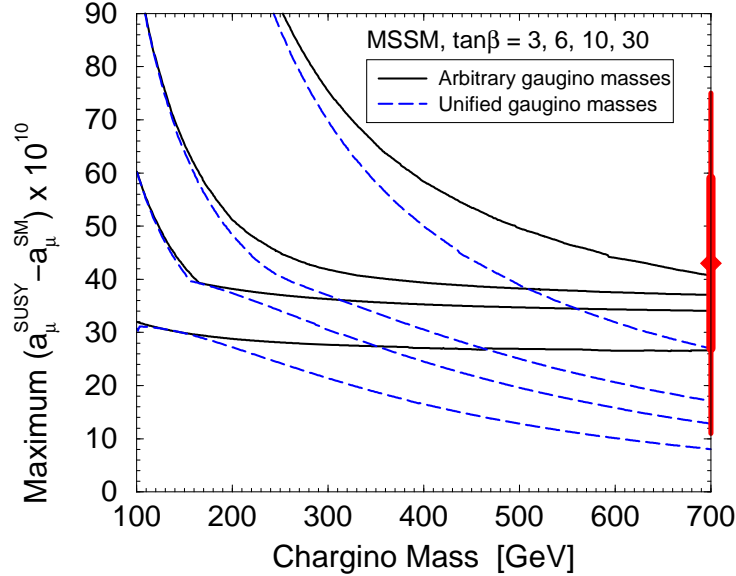}
\caption{Effects of a gaugino mass unification requirement: 
the maximum possible values for the
supersymmetric contribution to the muon anomalous magnetic 
moment, as a function of the lighter chargino mass. 
The solid lines are the general results as before
for $\tan\beta = 3,6,10,30$, while the dashed lines are obtained
with the additional condition $M_1 = (5/3) \tan^2\theta_W M_2$ imposed.
The maximum allowed value of $|\mu|$ is 1000 GeV.
All charged superpartners
are required to be heavier than 100 GeV, and the lightest
neutralino is required to be heavier than 50 GeV. 
The red bars on the right vertical
axis indicate the 1-$\sigma$ and 2-$\sigma$ allowed regions from
the Brookhaven E821 experiment. 
(The corresponding 
plot 
as a function of the lighter smuon mass is essentially
unaffected by the gaugino mass unification condition.)
\label{maxgeneralunif}}
\end{figure}

It is also interesting to see how our results would change if one
restricts to a class of models that make the usual
assumption of gaugino mass unification 
predicted by
supergravity-inspired models with unification of gauge couplings
and universal soft-supersymmetry breaking couplings,
namely
\beq
M_1 = {5\over 3} \tan^2\theta_W M_2\simeq 0.5 M_2 .
\label{gauginomassunification}
\eeq
It is plausible from a model-building perspective
that slepton and Higgs soft squared masses
can be affected by unknown $D$-term contributions 
\cite{Drees:1986vd}-\cite{Kolda:1996iw}
and other sources of
non-universality. This supports the idea of an unrestricted parameter space
for $m^2_{L}$, $m^2_{R}$,
$\mu$, and $A_\mu$, 
while still maintaining
the condition eq.~(\ref{gauginomassunification}).
Therefore we show the effects of imposing this assumption
on the parameter space in fig.~\ref{maxgeneralunif}.
This graph shows that
requiring gaugino mass unification does significantly
impact the maximum obtainable $a_\mu^{\rm SUSY} - a_\mu^{\rm SM}$ for
larger values of the chargino mass. This is clearly because if
gaugino mass unification is imposed, heavy charginos
necessarily means that the neutralino-smuon loop also decouples.
Without the gaugino mass unification requirement, a significant
contribution from the lightest smuon and bino-like neutralino loop can be
independent of the chargino masses. However, the results for
the maximum $a_\mu^{\rm SUSY} - a_\mu^{\rm SM}$ as a function of the
lighter smuon mass are essentially unaffected by the requirement of gaugino
mass unification, since the bounds in that case
are saturated by models with lighter charginos anyway.

\subsection{Constraints on the effects of complex phases}

\begin{figure}[tpb]
\dofigs{3.1in}{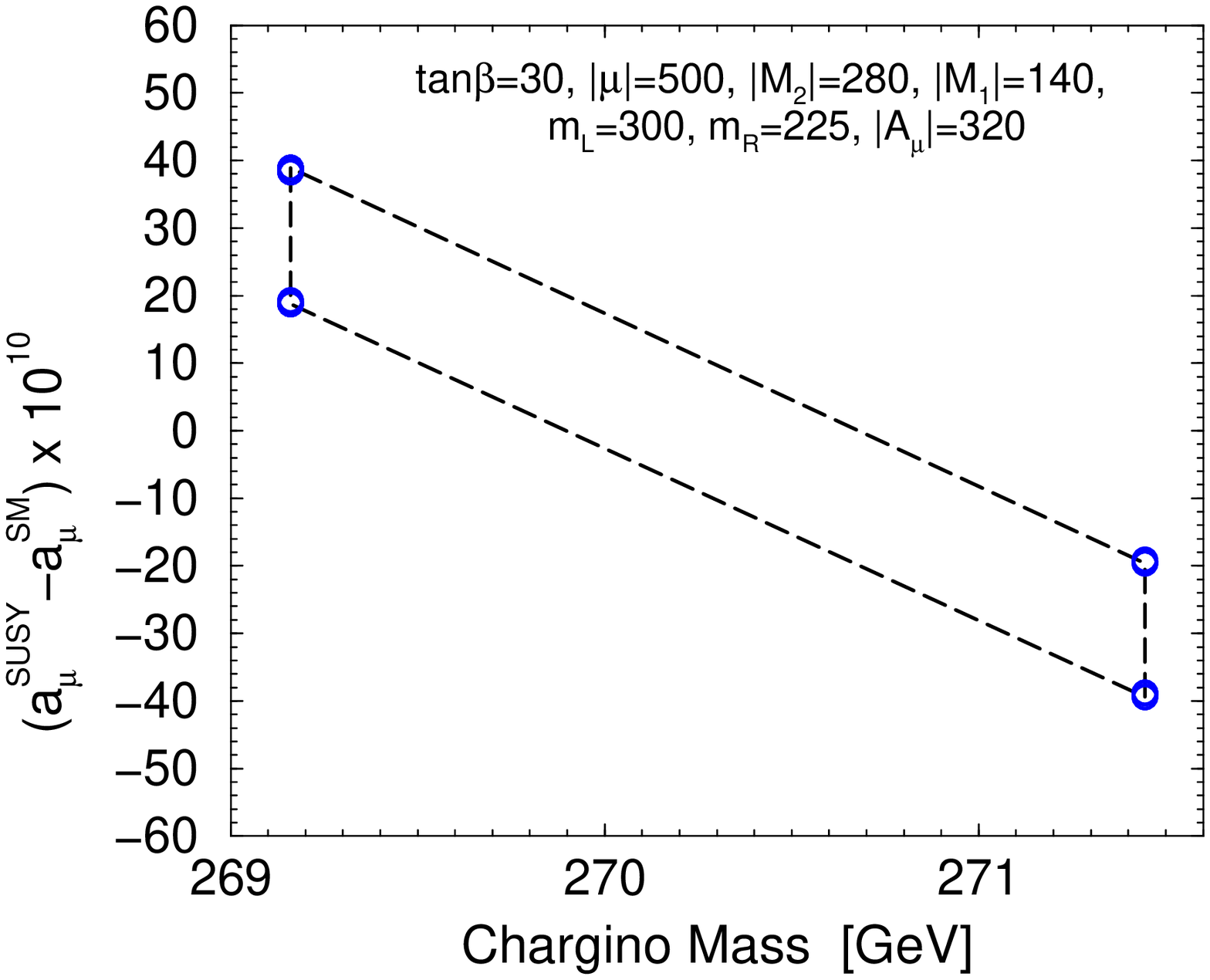}{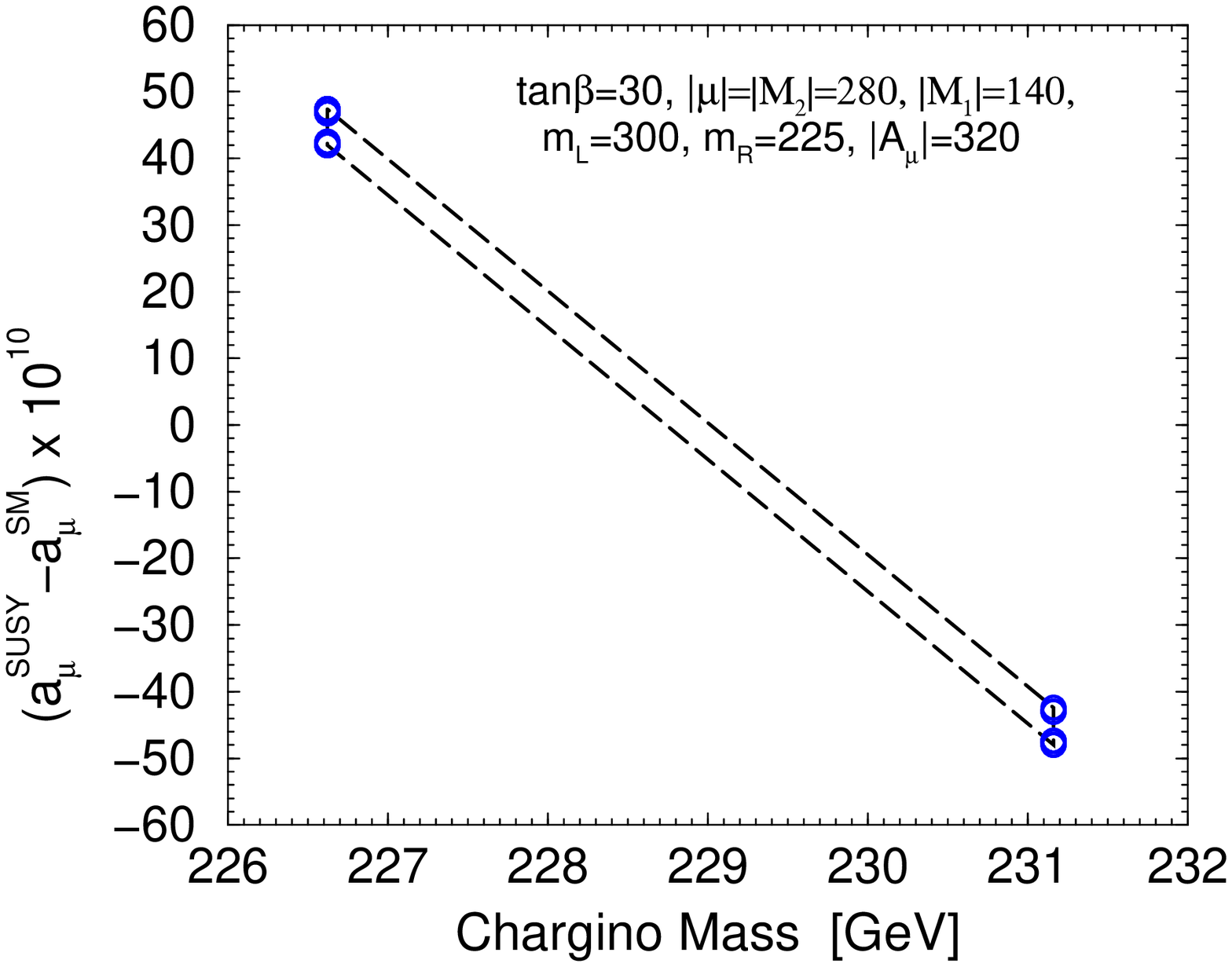}
\caption{Effects of phases: 
the contribution to the muon anomalous magnetic moment as a
function of the lighter chargino mass, with magnitudes of all parameters
held fixed. The dashed lines enclose the region obtained as the phases of
all parameters are varied over all possible values. The circles at the
corners of the regions are obtained when all parameters are
required to be real. The graph on the left is obtained for a model
close to a dilaton-dominated supergravity model, while the graph on the
right is the same but with $|\mu|$ adjusted to equal $|M_2|$.
\label{phases}}
\end{figure}

The above results were obtained for general values of all phases
of $M_1$, $M_2$, $\mu$, and $A_\mu$. It is useful to remark
that the maximum values of $a_\mu^{\rm SUSY} - a_\mu^{\rm SM}$, for fixed
magnitudes of the parameters, is generally
obtained when they are all real. For example, this is
illustrated for a particular choice of parameters (close to a
dilaton-dominated supergravity model) in fig.~\ref{phases}.
In the left graph, we show $a_\mu^{\rm SUSY} - a_\mu^{\rm SM}$ as a function of
chargino mass, with fixed $\tan\beta=30$, $|M_1| = 140$ GeV, $|M_2| = 
280 $ GeV, $|\mu| = 500$ GeV, and slepton parameters
$m_{L} = 300$ GeV,  
$m_{R} = 225$ GeV, and $|A_\mu |= 320$ GeV. 
The range of values for $\asusy$, 
obtained by varying over all possible phase values, fills out
the region
enclosed in the solid lines, while the circles at the corners denote
the points obtained when all parameters are real. In the right-hand graph,
the same thing is done for the same model, but with $|\mu| = $ 280 GeV
(equal to $M_2$) so that chargino mixing and neutralino mixing effects 
are larger. 

This illustrates
that while the dependence on the phases is quite strong as has been noted
in ref.~\cite{Ibrahim:2000hh,Ibrahim:2000aj}, the maximal
contribution to 
the muon anomalous magnetic moment occurs for real parameters;
in particular it usually occurs for positive 
real $\mu$, if $M_1$ and $M_2$ are both positive and real.  
This result is not surprising.  Unlike chiral violating interactions,
CP-violation breaks no symmetry critical to $a_\mu$ and so
its introduction cannot overwhelm the calculation.  The extremes of 
constructive and destructive interference naturally occur for
$e^{i\phi_k}=\pm 1$ (i.e., $\phi_k=0$ or $\pi$).
Therefore, imposing bounds from CP-violation experiments has no effect on
the results shown in figs.~1-3.

\section{Expectations in minimal supersymmetry models}
\setcounter{equation}{0}
\setcounter{footnote}{1}

The full supersymmetry parameter space, including all supersymmetric
masses and mixing angles, contains well over 100 free parameters.  The
vast majority of this parameter space is ruled out by experimental
measurements of proton lifetime, flavor changing neutral currents, and
CP-violating observables.  Ideas to solve these problems in supersymmetry
are varied.  However, there exists two baseline, or minimal models, that
are largely immune from all past experimental constraints, and are
often employed to estimate accessibility of supersymmetry in new
experiments.  These two models are called ``minimal supergravity'' 
(SUGRA) and ``minimal gauge mediation'' (GMSB).

One advantage of having minimal models as baselines for comparing
expectations of supersymmetry is that they existed and were well-motivated
before anomalies were seen by experiment. They therefore
provide a more restrictive but still dispassionate view of how easily
supersymmetry can accommodate 
non-SM effects, complementary to that obtained from the fully
model-independent
framework described in the previous section.
When the measurement of 
$R_b=\Gamma (Z\to b\bar b)/\Gamma(Z\to {\rm had})$ appeared to have
a $3\sigma$ deviation from the SM prediction, it was shown that 
SUGRA could not accommodate it~\cite{Wells:1994cu}.  One could attain
$R^{\rm expt}_b$ in supersymmetry only by entertaining unusual
corners of parameter space.  It might be accurate to say that 
the SUGRA analysis of $R_b$ has turned out to be the most enlightening
one.  

We perform the SUGRA analysis here for similar dispassionate
reasons.  One expects a large class of viable supersymmetric theories
to be in the neighborhood of SUGRA, especially for the subset of
MSSM parameters that enter into the $a_\mu$ calculation.  We also
do an analysis for GMSB since that constitutes a separate, equally
interesting minimal model positioned in a different large neighborhood of 
viable supersymmetric theories.  For the reader's ability to reproduce
our results, we define our models by feeding 
SUGRA and GMSB spectra from the ISAJET
sugrun code~\cite{Baer:1999sp} into the $a_\mu$ formulas presented above.  

\subsection{Minimal supergravity}

SUGRA simplifies the derivation of the superpartner spectrum by assuming
that all gauginos unify at the grand unified (GUT) scale with mass
$m_{1/2}$, and all scalars unify at the GUT scale with mass $m_0$.
Additional free parameters are $\tan\beta$ (the ratio of Higgs vacuum
expectation values), $A_0$ (common trilinear scalar coupling at the GUT
scale), and the sign of $\mu$ (the superpotential
Higgs mixing mass parameter with sign convention of 
Refs.~\cite{Haber:1985rc,Martin:1997ns}).  
For a more thorough description of SUGRA and this parameterization, 
see~\cite{Kane:1994td,Abel:2000vs}.

\begin{figure}[tpb]
\dofigs{3.5in}{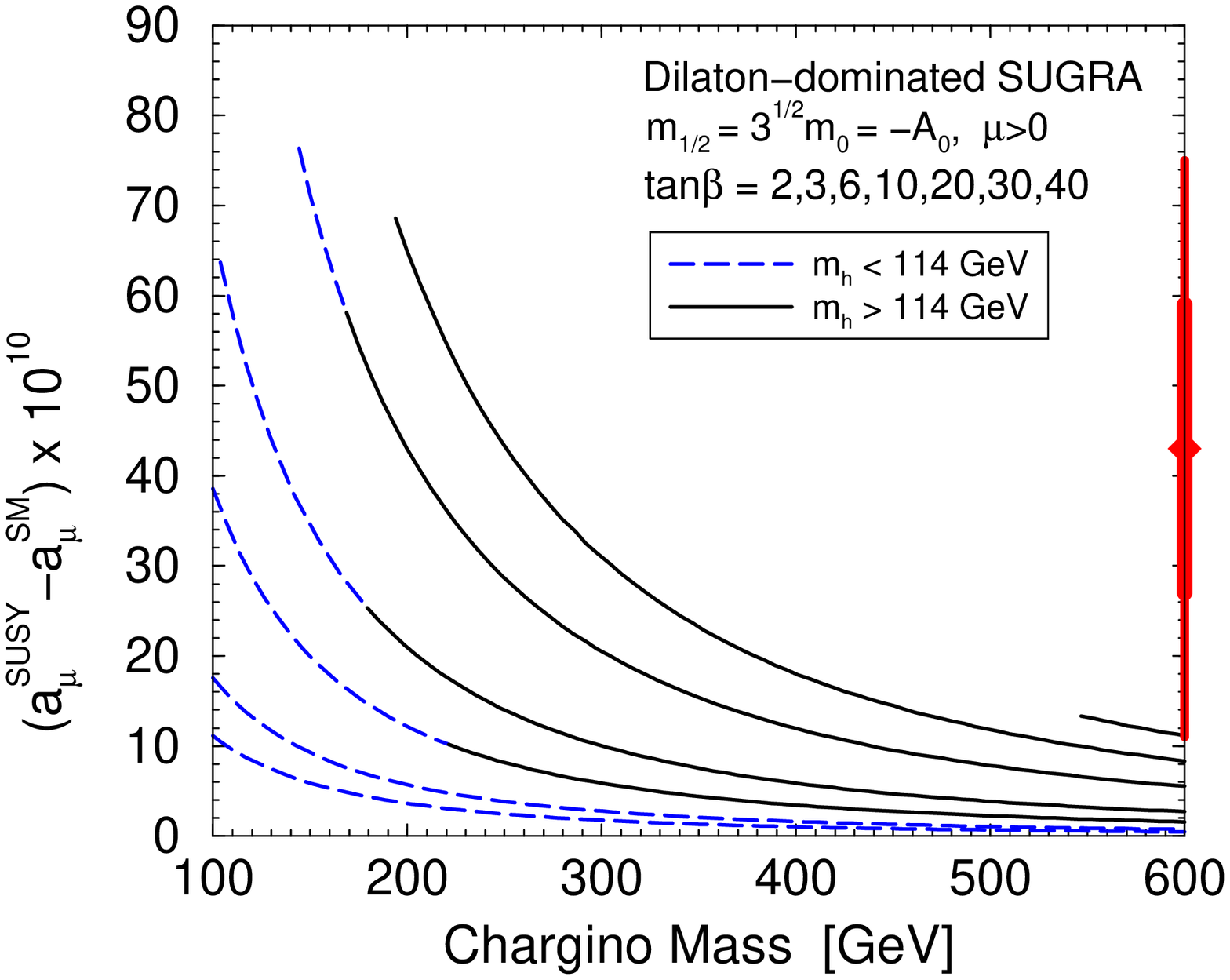}{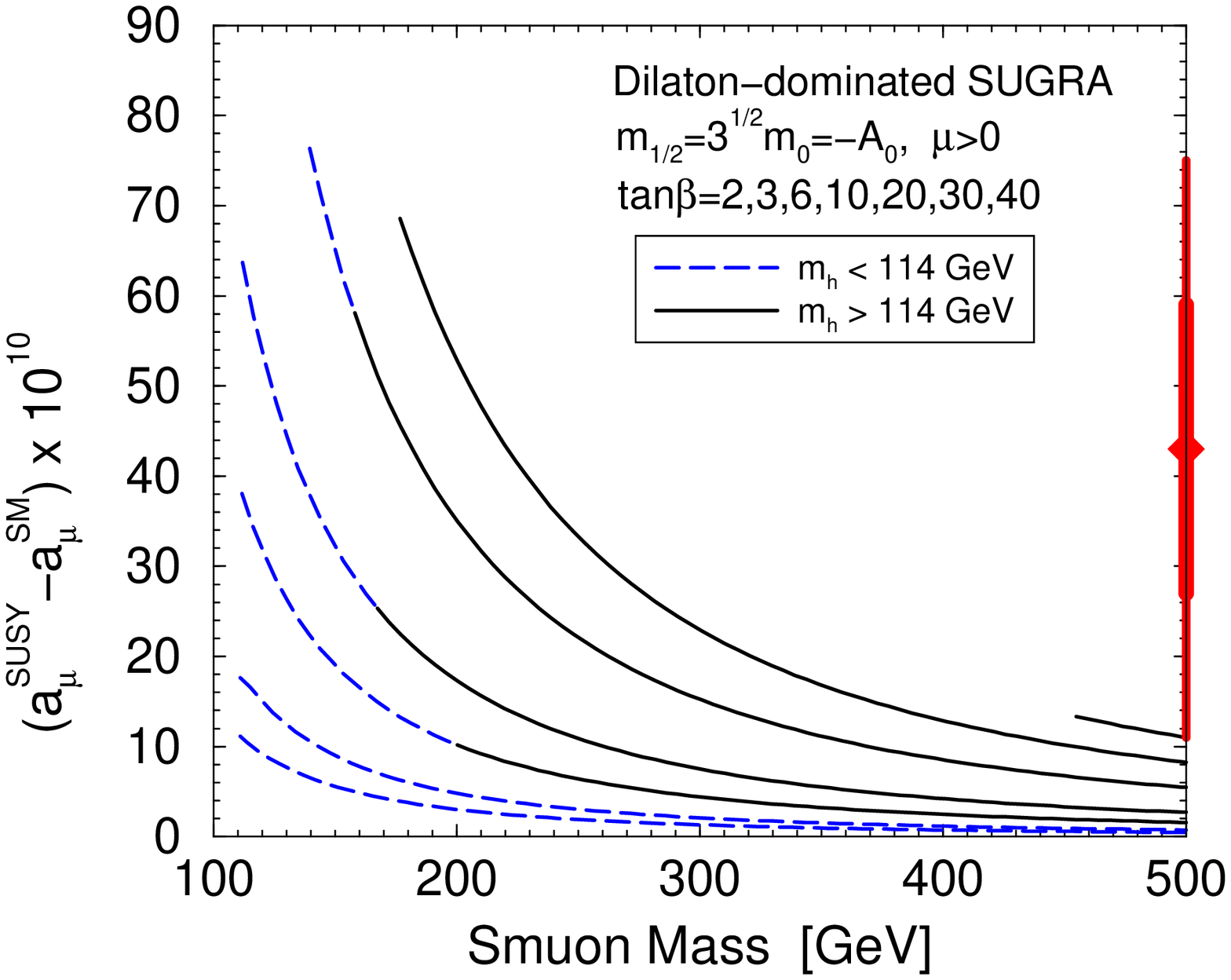}
\caption{Predictions for $a_\mu^{\rm SUSY} - a_\mu^{\rm SM}$ in
dilaton-dominated supergravity models 
with various $\tan\beta=2,3,6,10,20,30,40$ (from bottom to top), 
as a function
of the lighter chargino mass and the lighter smuon mass. 
All charged superpartners are required to have mass above $100\gev$.
The solid
lines indicate where the lightest Higgs scalar boson mass $m_h$ exceeds
its approximate LEP2 bound of 114 GeV, while the dashed lines indicate
where $m_h < 114 $ GeV. The red bars on the right vertical
axes indicate the 1-$\sigma$ and 2-$\sigma$ allowed regions from
the Brookhaven E821 experiment.
\label{dilaton}}
\end{figure}

We will illustrate the generic effects that SUGRA has on $a_\mu$ by
initially restricting ourselves to the so-called dilaton dominated
scenario where
\bea
m_{1/2}=-A_0=\sqrt{3}m_0 ~~~~{\rm (dilaton~dominated)}.
\eea
In fig.~\ref{dilaton}
we have plotted $\delta a^{\rm SUSY}_\mu$ vs.\ superpartner mass (chargino 
and lightest smuon) for various $\tan\beta$.  The dashed lines mean
$m_h<114\gev$, in apparent conflict with LEP2 bounds on the Higgs 
boson~\cite{Abreu:2001fw}.  
If $m_h\simeq 114\gev$ turns out to be the actual Higgs boson mass, as some 
tantalizing data seem to suggest, then one can spot the prediction for
$\delta a^{\rm SUSY}_\mu$ by focusing on the interface between the
dashed lines and the solid lines.

Going from right to left, some of the lines terminate abruptly.  The reason
for this is that we have required all lines,
dashed or solid, to be consistent with $m_{\tilde \tau_1}>100\gev$, which is
our conservative cut based on anticipated limits from the
final LEP2 analyses. Since the
stau mass matrix in most models, including this one, is correlated closely
with the smuon mass matrix, we can test unambiguously
if $m_{\tilde \tau_1}<100\gev$.
When $\tan\beta$ is large,
$m_{\tilde\tau_1}<m_{\tilde\mu_1}$ because
the off-diagonal part of the mixing matrix, $-m_{\mu,\tau}\mu \tan\beta$, is
larger for the $\tilde\tau$ than $\tilde\mu$, and level repulsion of mass
eigenstates will push $m_{\tilde\tau_1}$ lower than $m_{\tilde\mu_1}$.
The available smuon masses are also constrained by $m_{\chi_1^\pm}
> 100$ GeV for $\tan\beta=2,3,6$.
For these reasons, fig.~\ref{dilaton} has some lines ending within the
plots.

As expected, the higher values of $\tan\beta$ have higher $\asusy$
contributions, have less problem with the $m_h>114\gev$ constraint, and have
more problem with the $m_{\tilde\tau_1}>100\gev$ constraint.  The
Higgs and $\tilde\tau_1$ mass constraints are competing effects in the
drive to get high $\asusy$.  In the end, large $\tan\beta$ still wins
out and we can easily get within the $1\sigma$ allowed region by requiring
$\tan\beta\gsim 20$, $m_{\chi^\pm_1}\lsim 260\gev$ and 
$m_{\tilde\mu_1}\lsim 230\gev$; or
$\tan\beta\gsim 30$, $m_{\chi^\pm_1}\lsim 325\gev$ and 
$m_{\tilde\mu_1}\lsim 280\gev$.

\begin{figure}[tpb]
\dofigs{3.5in}{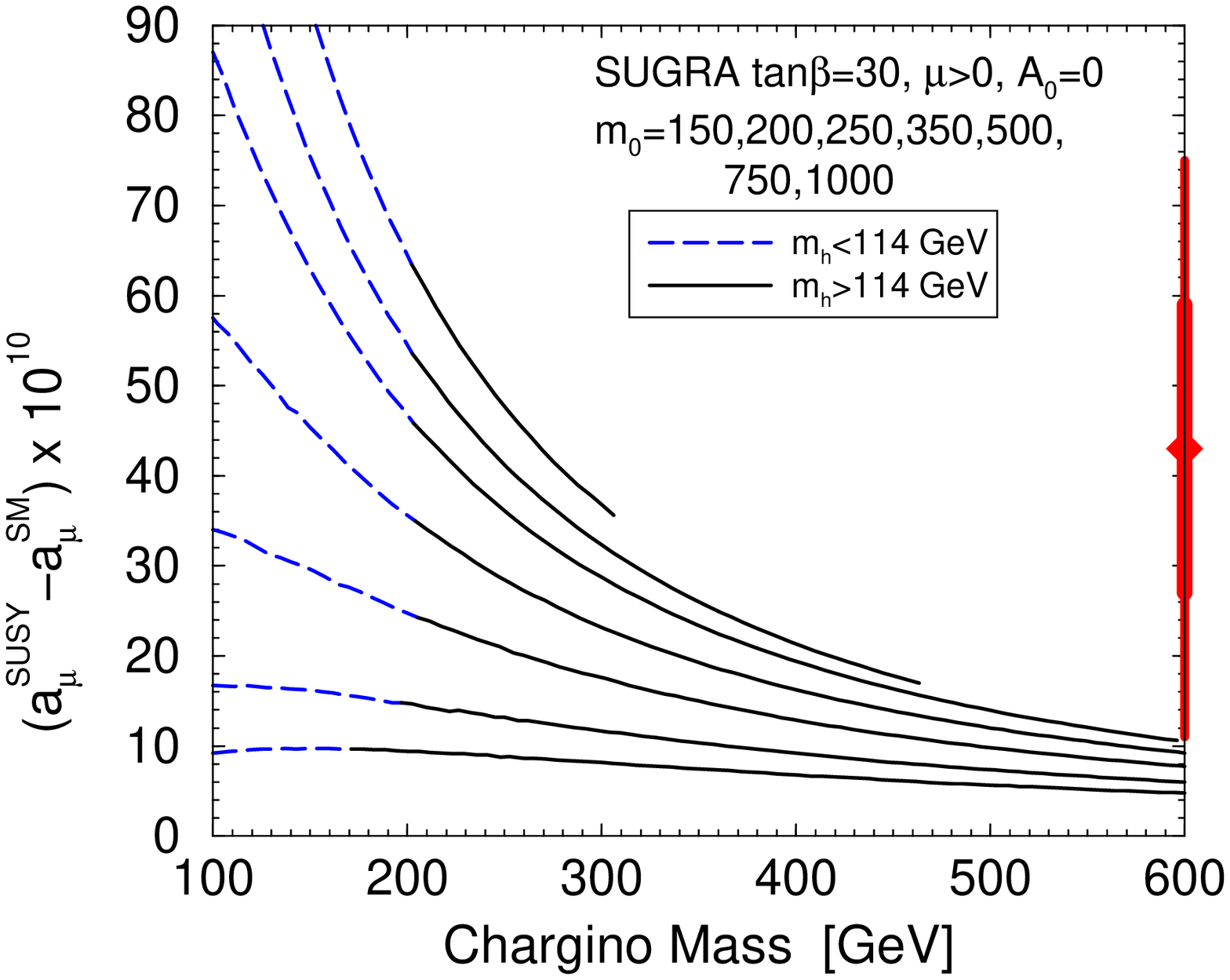}{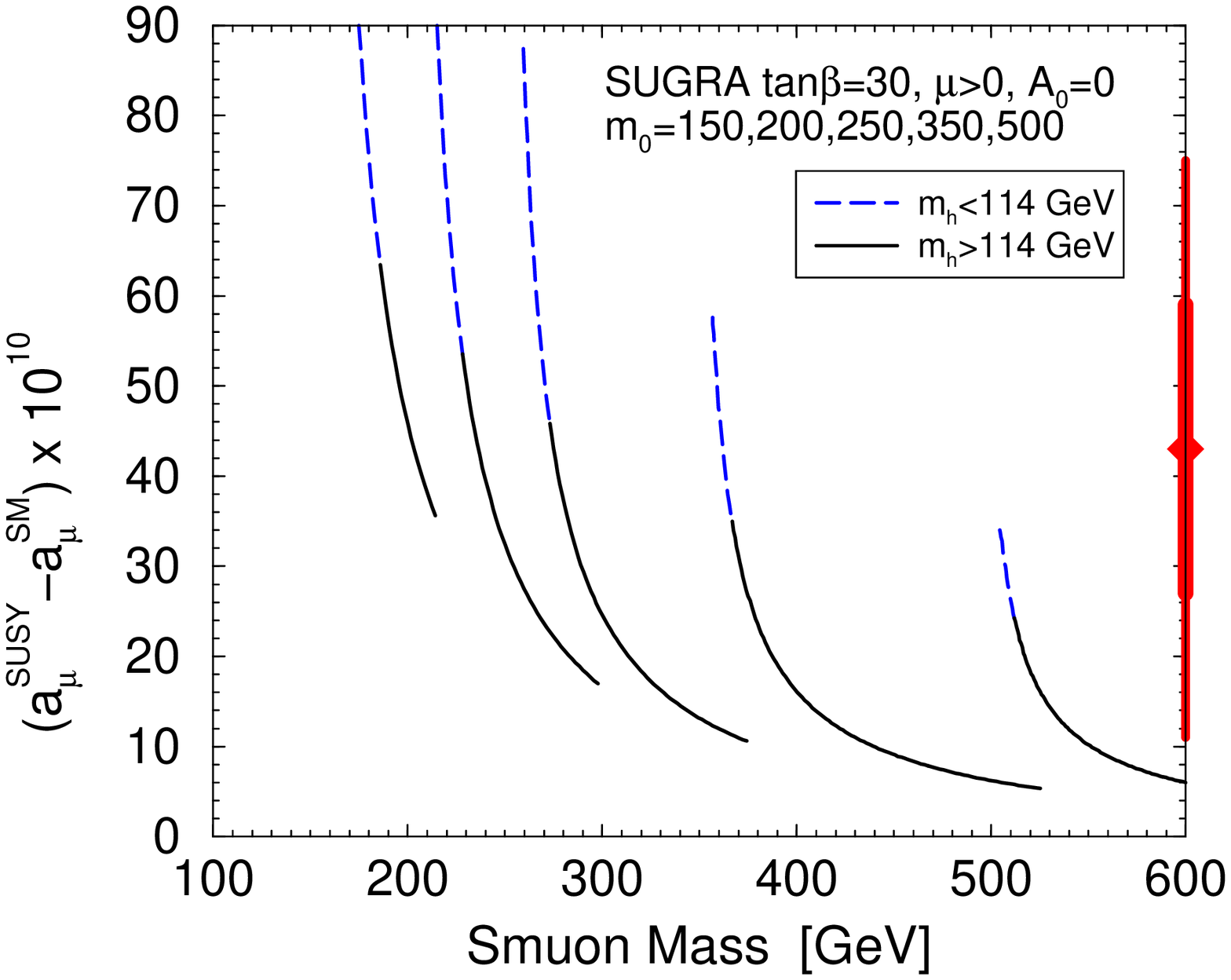}
\caption{Predictions for $a_\mu^{\rm SUSY} - a_\mu^{\rm SM}$ in
minimal supergravity models 
with various $m_0=150, 200, 250, 350, 500, 750, 1000\gev$, from
top to bottom as a function
of the lighter chargino mass and from left to right as a function of
the lighter smuon mass. 
All charged superpartners are required to have mass above $100\gev$.
The solid
lines indicate where the lightest Higgs scalar boson mass $m_h$ exceeds
its approximate LEP2 bound of 114 GeV, while the dashed lines indicate
where $m_h < 114\gev$.
The red bars on the right vertical
axes indicate the 1-$\sigma$ and 2-$\sigma$ allowed regions from
the Brookhaven E821 experiment.
\label{sugra m0}}
\end{figure}

Since $\asusy$ scales as $\tan\beta$ for large $\tan\beta$ we now
suppress discussion of this 
known behavior by fixing $\tan\beta=30$ and vary $m_0$ within
the SUGRA framework.  
We are comfortable with this larger $\tan\beta$  
choice
for
another reason.  Namely, $t-b-\tau$ Yukawa coupling unification
is most easily satisfied for larger $\tan\beta$ 
theories~\cite{Hall:1994gn}.  This tri-unification
of Yukawa couplings is preferred in minimal version of $SO(10)$ grand
unification. Fig.~\ref{sugra m0} plots the prediction of $\asusy$
vs.\ chargino mass and lightest slepton mass for various values of
$m_0$.  Again, the dashed lines
indicate $m_h<114\gev$.  
The dashed lines terminate on the left where $m_{\chi_1^\pm} < 100$
GeV.
Going from left to right, the solid lines terminate 
because ${\tilde \tau_1}$ becomes the LSP.  There are two problems with this.
First, charged LSPs are cosmologically
disfavored~\cite{DeRujula:1990fe}.
And second, even if one assumes $R$-parity violation will decay
away the dangerous charged relics, we would have to give up on the
very attractive neutralino LSP of SUGRA.  For this reason we have terminated
the lines when $m_{\tilde \tau_1}<m_{\chi^0_1}$, although it is easy enough
to visually follow where the lines would have extended in the higher
chargino mass region.

From fig.~\ref{sugra m0} we learn that for a large value of $\tan\beta$,
such as the choice here of 30, large contributions are possible for
$\asusy$, but the superpartner effects decouple rapidly. 
For $\tan\beta=30$, one requires $m_{\chi^\pm_1}\lsim 350\gev$
and $m_{\tilde\mu_1}\lsim 500\gev$ to be within $1\sigma$ of the
measured value.  Both these masses increase to approximately $600\gev$
to find oneself within $2\sigma$ of the measured value.

In short, the SUGRA model with large $\tan\beta$ generically gives
large values of $\asusy$ for superpartners with mass at least
as high as three times the
current experimental limits.  Therefore, SUGRA or some approximate to
it would not be a surprising solution to the measured non-SM contribution
of the muon anomalous magnetic moment.

\subsection{Minimal gauge mediation}

GMSB organizes the superpartner spectrum in an entirely different way,
but with equal simplicity, by assuming that all superpartners get their
masses by interacting through ordinary gauge bosons with messenger
fields that feel supersymmetry breaking.  In the minimal model the messenger
fields are assumed to be equivalent to an integer number ($N_5$)
of complete multiplets of ${\bf 5}+{\bf \bar 5}$
fields of $SU(5)$.  Along with $N_5$, other free parameters are the
supersymmetry breaking scale $\sqrt{F}$, the messenger mass scale
$M_m$, and $\tan\beta$.  For simplicity in this analysis 
we assume the reasonable
relation $M_m=100 \Lambda$, where $\Lambda = F/M_m$ sets the
scale of the MSSM sparticle masses.  
For a more thorough description of 
GMSB and this parameterization, see~\cite{Giudice:1999bp,Culbertson:2000am}.

\begin{figure}[tpb]
\dofigs{3.5in}{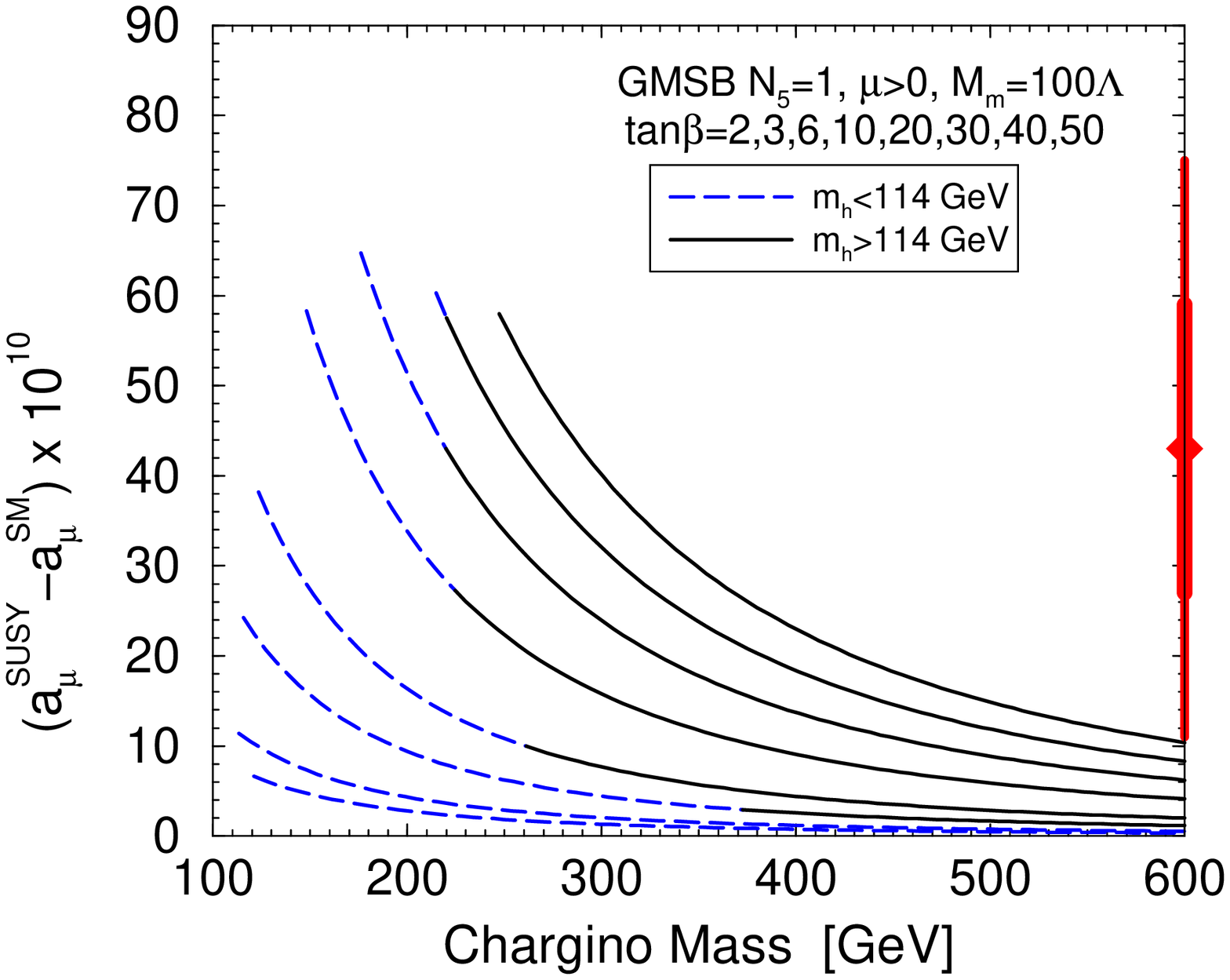}{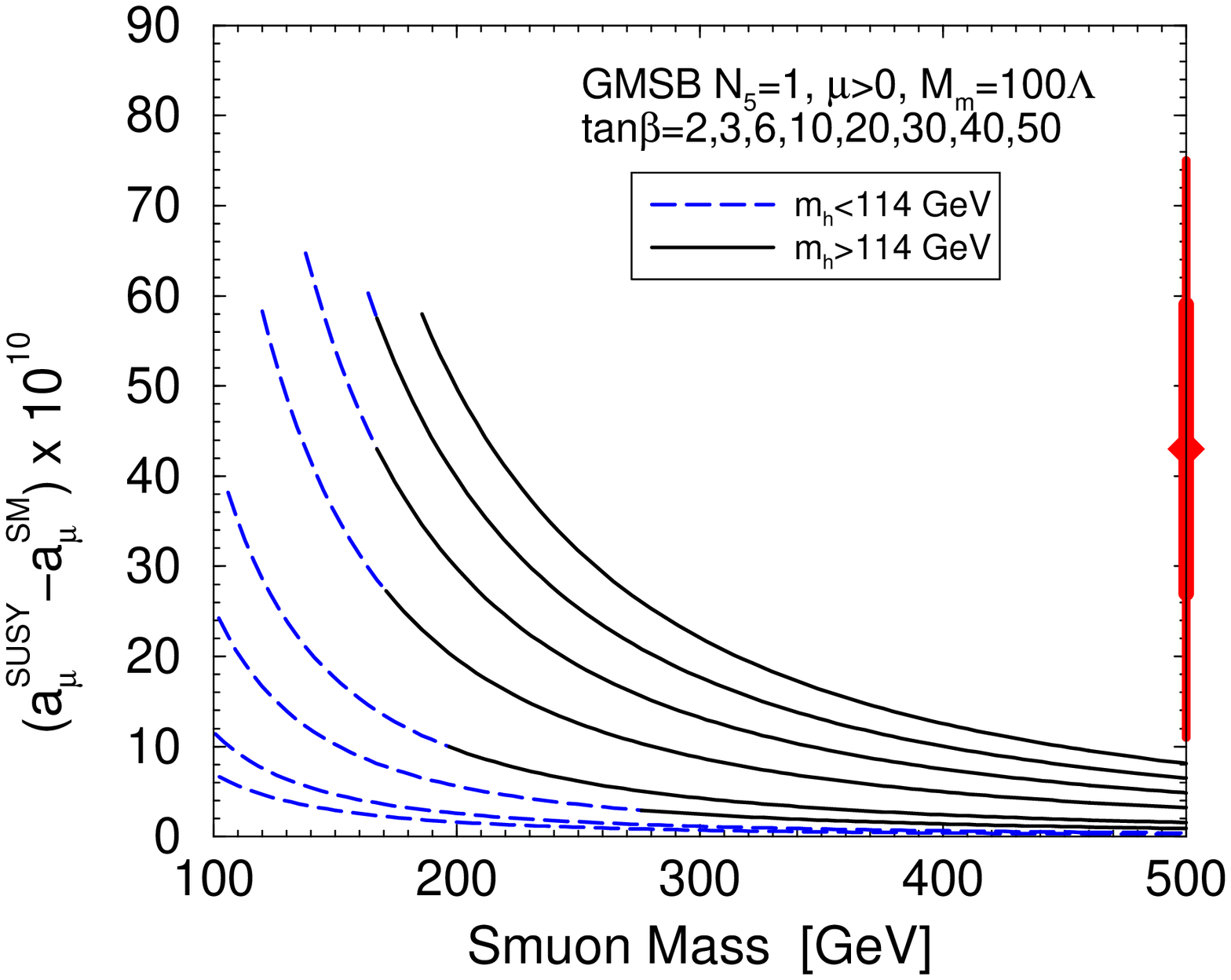}
\caption{Predictions for $a_\mu^{\rm SUSY} - a_\mu^{\rm SM}$ in
minimal GMSB models for $N_{5+\bar 5}=1$ and various 
$\tan\beta=2$, 3, 6, 10, 20, 30, 40 and 50 (from bottom to top),
as a function
of the lighter chargino mass and the lighter smuon mass.
All charged superpartners are required to have mass above $100\gev$.
The solid
lines indicate where the lightest Higgs scalar boson mass $m_h$ exceeds
its approximate LEP2 bound of 114 GeV, while the dashed lines indicate
where $m_h < 114\gev$. The red bars on the right vertical
axes indicate the 1-$\sigma$ and 2-$\sigma$ allowed regions from
the Brookhaven E821 experiment.
\label{gmsb tanbeta}}
\end{figure}

Our first illustration of the GMSB predictions will be for the
most minimal model of one messenger ${\bf 5}+{\bf \bar 5}$, i.e.\ $N_5=1$.
In fig.~\ref{gmsb tanbeta} we plot $\asusy$ vs.\ lightest chargino mass
and
lightest smuon mass for various $\tan\beta$.
Again, the dashed lines represent $m_h<114\gev$ for comparison with
LEP2 searches, and the lines terminate to the left because
$m_{\tilde\tau_1}<100\gev$.

\begin{figure}[tpb]
\dofigs{3.5in}{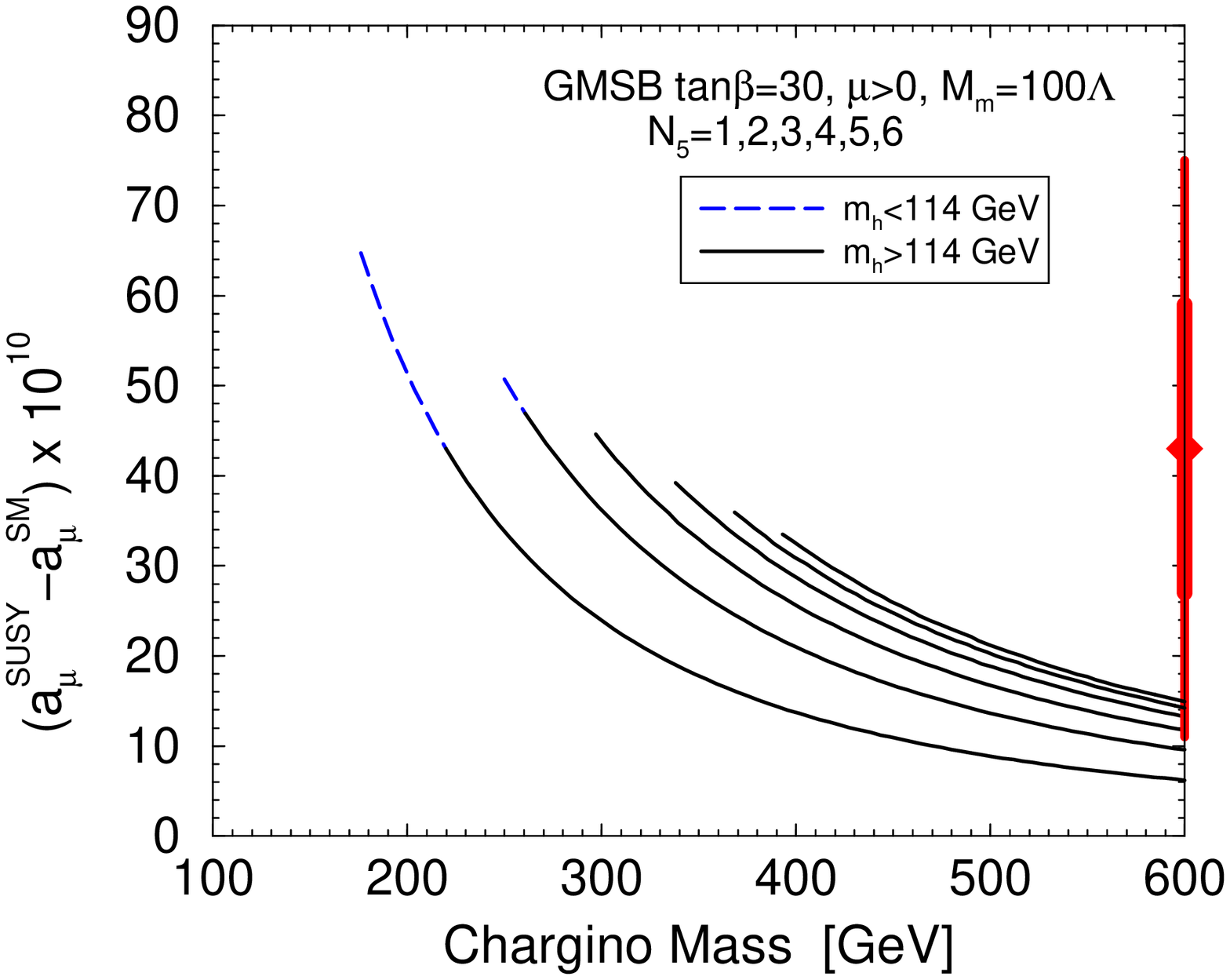}{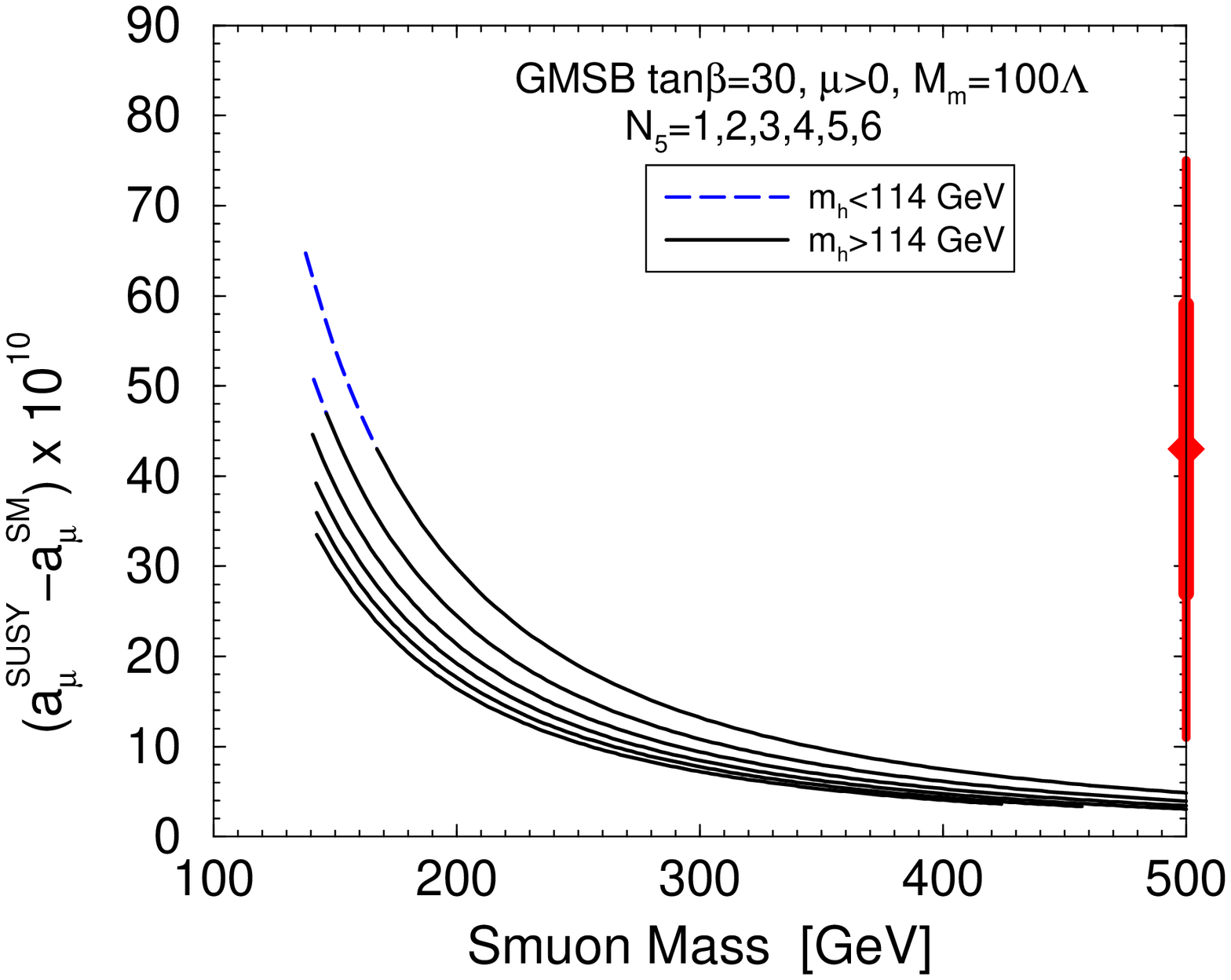}
\caption{
Predictions for $a_\mu^{\rm SUSY} - a_\mu^{\rm SM}$ in
minimal GMSB models for $\tan\beta=30$ and various $N_5=1$, 2, 3, 4, 5, 6
from bottom to top
as a function
of the lighter chargino mass, and from top to bottom
as a function of the lighter smuon mass.
All charged superpartners are required to have mass above $100\gev$.
The solid
lines indicate where the lightest Higgs scalar boson mass $m_h$ exceeds
its approximate LEP2 bound of 114 GeV, while the dashed lines indicate
where $m_h < 114\gev$. The red bars on the right vertical
axes indicate the 1-$\sigma$ and 2-$\sigma$ allowed regions from
the Brookhaven E821 experiment.
\label{gmsb N5}}
\end{figure}

We witness from fig.~\ref{gmsb N5} yet another example of how large
$\tan\beta$ enhances the value of $\asusy$.  For large but reasonable 
values of $\tan\beta$, $\asusy$ is within
$1\sigma$ of the measured value.  Again, masses can be several times 
heavier than the current limits to accomplish this, and no additional
constraints such as $m_h$ or $m_{\tilde\tau_1}$ limits disturb the result.
An intriguing feature
of this plot is the near-equal predictions of SUGRA dilaton
dominated scenario and $N_5=1$ GMSB for fixed chargino mass.  This only
means that in both these
minimal models the relative masses of the charginos and smuons are
close for the same values of $\tan\beta$.  

Variations in the spectrum occur for different values of $N_5$. In 
fig.~\ref{gmsb N5}, we fix $\tan\beta=30$ and plot $\asusy$ for 
various $N_5$.  The higher the number of ${\bf 5}+{\bf \bar 5}$
representations the higher $\asusy$ for a given chargino mass.
This is simply because $m_{\tilde\tau_1}/m_{\chi^\pm_1}\propto 1/\sqrt{N_5}$.
Of course, if $m_{\tilde\tau_1}$ dips below $100\gev$ the line 
is not extended,
which explains the curious result in fig.~\ref{gmsb N5} that
the highest allowed $\asusy$ for a fixed $\tan\beta$ comes from
{\it lower} $N_5$.

In short, the simplest GMSB models
have similar predictions as SUGRA for $\asusy$, and
can naturally produce a result within $1\sigma$ of the measured value
for reasonable superpartner masses 
well above direct experimental mass limits.  We find the
results for SUGRA and GMSB encouraging for the supersymmetric 
interpretation of $a_\mu$.

\section{Discussion of correlating phenomena}
\setcounter{equation}{0}
\setcounter{footnote}{1}

The anomalous magnetic moment of the muon is just one observable out
of many that supersymmetry can affect.  Using just this one quantity
to divine predictions for other observables is difficult for the obvious
reason that each observable requires a different set of supersymmetry
masses and mixing angles.  Furthermore, even within a narrowly
defined version of supersymmetry, such as SUGRA, a single value
of $\asusy$ maps to a vast parameter space within the model.  

\subsection{Superpartners at colliders}

With the above caveats we make a few general comments on expected 
correlating phenomena.  All remarks are based on one generally 
drawn conclusion:
the large $a_\mu$ measurement at E821 likes a supersymmetric
interpretation with larger $\tan\beta$ and
lighter superpartners.  The exact values of $\tan\beta$ and superpartner
masses are model-dependent.  In the previous sections, 
we have shown that in the most general
MSSM no meaningful bounds can be placed on the chargino mass, and weak
bounds can be placed on the lightest smuon mass. For example, if
$\tan\beta <20$ then $m_{\tilde \mu_1}\lsim 500\gev$.  However, we
readily admit that one 
does not find {\it generically} in MSSM parameter space
that the lightest smuon mass can be above several hundred
GeV and $\asusy$ within $1\sigma$.  
This assessment is made by analyzing minimal models and making
agenda-less tours in supersymmetry parameter space.  
Therefore, we would {\it cautiously} 
agree~\cite{Everett:2001tq,Feng:2001tr,Komine:2001fz} that the
first statement to make about correlating phenomena is that smuons
should be light.  Smuons are notoriously difficult objects to discover
at hadron colliders~\cite{Baer:1994ew}. They are relatively easy to find
at $e^+e^-$ machines, but of course the center of mass energy must
be sufficient to produce them.

\subsection{Higgs boson mass}

Other conclusions are a bit more subtle.  For example, large $\tan\beta$
is also preferred by Higgs search results at LEP2.  Higgs bosons greater
than
(or equal to) $114\gev$ have put strain on low $\tan\beta$ 
models (see, e.g.\ \cite{Kane:2000kc,Ellis:2000sv}).
This can be seen most readily by the prediction for the lightest
supersymmetric scalar Higgs boson mass eigenstate in the
$m_Z^2/m_A^2\ll 1$ limit:
\bea
m^2_h=m_Z^2\cos^2\, 2\beta + \frac{3 g_2^2m_t^4}{8\pi^2m_W^2}\ln 
      \frac{\Delta^2}{m_t^2}.
\label{higgs mass}
\eea
This formula is exact if one is willing to tolerate an extremely
complicated form for $\Delta^2$, otherwise it can be interpreted
as approximate~\cite{Ellis:1991nz,Haber:1991aw}
with $\Delta^2\simeq m_{\tilde t_1}m_{\tilde t_2}$.
The first term representing the tree-level mass prediction grows
larger with $\tan\beta$.  Furthermore, although not obvious from
the form of eq.~(\ref{higgs mass}), the radiative corrections to the
Higgs mass can increase with larger $\tan\beta$ also (i.e., $\Delta$ depends
mildly on $\tan\beta$).
As a simple illustration, for $\Delta=1\tev$ ($=500\gev$) 
we need $\tan\beta\geq 2.8$ ($\geq 12$) to
ensure $m_h>114\gev$.  This tendency for higher $\tan\beta$ with lower
superpartner masses to satisfy the Higgs mass bound is in
the same direction as the requirements of $\asusy$.

We remark that we would not be surprised if the Higgs sector were 
different than the simple two Higgs doublets of the MSSM.  For example,
an additional singlet field with superpotential term $\lambda SH_uH_d$
may even be more preferred since it can lead to spontaneous generation
of the $\mu=\lambda\langle S\rangle$ term, among other advantages.
The lightest Higgs boson in this case would then get a contribution
to its mass proportional to $\lambda^2 v^2$, potentially
making $\tan\beta$ limits from Higgs boson mass in the MSSM irrelevant,
depending on the size of the Yukawa coupling $\lambda$.  

\subsection{Neutralino dark matter}

Another potentially important correlation is in dark matter relic
abundance and dark matter detection.  Several 
authors~\cite{Drees:2000bs,Baltz:2001ts,Chattopadhyay:2001vx,Arnowitt} 
have noted
that within some specific frameworks, e.g.\ SUGRA, the dark matter
detection rate prediction is large when $\asusy$ is large.
This is partly because coherent scattering of dark matter
off nuclei similarly requires a chirality flip and so 
is enhanced by larger $\tan\beta$.   

Recently, DAMA
has claimed a signal in the annual modulation of WIMP-nuclei 
scattering~\cite{Bernabei:2000qi}.
The supersymmetric interpretation implies a large 
spin-independent coherent scattering cross-section, which is
easier to attain at large $\tan\beta$~\cite{Accomando:2000eg}.  
The DAMA signal
may or may not be real, but the correlation remains: large $\tan\beta$
implied by $\asusy$ generally implies larger scattering cross-sections
for dark matter detectors.  Of course, it is possible that supersymmetry
has nothing to do with dark matter because $R$-parity
is not conserved, or some other reason, in which case these issues
become irrelevant.  

As for relic abundance of the lightest neutralino, light sleptons could
create a problem for the supersymmetric interpretation of dark matter 
since they induce an efficient $t$-channel annihilation
channel in $\chi^0_1\chi^0_1\to l^-l^+$.  The larger the annihilation
channel the smaller the relic abundance ($\Omega \sim 1/\sigma v$).
However, there is a large region of parameter space for 
light sleptons~\cite{Wells:1998ci,Ellis:2001ms}
(but heavier than $100\gev$) that is consistent 
with adequate thermal relic abundance to be cosmologically interesting,
$0.1\lsim \Omega h^2 \lsim 0.4$.  Furthermore, in the regions where
there is small thermal relic abundance from light sleptons or 
large coannihilation effects~\cite{Ellis:2000mm}, 
there are  non-thermal 
sources~\cite{Gherghetta:1999sw,Moroi:2000zb,Jeannerot:1999yn}
of the LSPs that could regenerate them as dark matter.  Therefore,
we do not think relic abundance considerations add significantly
to the dialog on $\asusy$ at this point.

\subsection{$B(b\rightarrow s\gamma)$ constraint}

Lastly, we remark on $B(b\to s\gamma)$.  There is a close similarity
between $a_\mu$ and $B(b\to s\gamma)$ in that both get large $\tan\beta$
enhancements from a higgsino-sfermion-fermion interaction vertex
with a down-fermion Yukawa coupling.  If the E821 experiment
had measured $\asusy \simeq -43\times 10^{-10}$ instead of
$+43\times  10^{-10}$, 
the measurement of
$B(b\to s\gamma)$ would have disfavored many supersymmetric
 interpretations.
However, it
happens that
$\asusy$ prefers $\mu >0$ (for real positive gaugino masses)
and large $\tan\beta$, 
and $B(b\to s\gamma)$ severely restricts $\mu < 0$ and large $\tan\beta$,
but does not significantly restrict $\mu >0$.

It is well-known that $M_3\mu >0$ is not as restricted by
B($b\to s\gamma)$~\cite{Baer:1998jq,Carena:2001uj,Ellis:2001ms} as 
$M_3\mu <0$, 
since the signs of the amplitudes
in this circumstance imply partial cancellations.
Recently, this conclusion was strengthened even more by the
evaluation of higher-order
calculations to $B(b\to s\gamma)$.  At higher order one must self-consistently
take into account the finite $b$-quark mass corrections which are enhanced
dramatically at large $\tan\beta$.  These corrections imply smaller
$b$-quark Yukawa coupling and therefore smaller magnitude for the
higgsino-squark-quark chirality flip.  From fig.~2
of~\cite{Degrassi:2000qf}
one can see the reduction in the supersymmetric prediction for large
$\tan\beta$ with $M_3 \mu >0$, rendering $B(b\to s\gamma)$ unable to
significantly
constrain large {\it positive} $\asusy$ scenarios.  

The above discussion
is mostly based on SUGRA-like relations among superpartner
masses.  A similar conclusion can be inferred from
ref.~\cite{Ellis:2001yu},
wherein $B(b\to s\gamma)$ has little 
impact on the viability of the CMSSM to
explain $\asusy$.  
The same discussion holds for GMSB since the squarks are even
heavier in that model, and $B(b\to s\gamma)$ was never much of a serious
constraint~\cite{Dimopoulos:1997yq} when all the uncertainties are
accounted for.  
Other theories of supersymmetry breaking such as anomaly mediated
supersymmetry breaking (AMSB)~\cite{Randall:1999uk,Giudice:1998xp} appear
to have
difficulty accomodating $\asusy$~\cite{Feng:2001tr,Choi}.  This difficulty
arises because $M_3 <0$ and the lightest gauginos are winos with small
positive $M_2$, leading to a
severe constraint on $\mu >0$ parameter space from $B(b\to s\gamma)$. The
$SU(5)$ model with a supersymmetry breaking F-term in the {\bf 24}
representation discussed at the end of sec.\ (1.2) also has $M_3 <0$.
However, in this case the ratios $|M_2/M_1|\simeq 6$ and $|M_3/M_1|\simeq
12$ imply that the lightest neutralino and smuon entering the $\asusy$
loop corrections would be significantly lighter than the squarks and
charginos that affect $B(b\to s\gamma)$.  Even though these mass
hierarchies make $B(b\to s\gamma)$ less important of a constraint than it
is in AMSB, careful evaluation of the next-to-leading order $B(b\to
s\gamma)$ prediction would need to be compared with experiment to
ultimately judge the viability of this model to explain $\asusy$.

One also must approach the $B(b\to s\gamma)$ observable
with a bit of caution when trying to rule out parameter
space consistent with $\asusy$.  Most analyses implicitly
assume that the theory prediction is precise, and it
need only fit into the range
obtained from experimental measurement, often quoted
to be between $(2-4)\times 10^{-4}$.  
The SM theory prediction~\cite{Kagan:1999ym} is
\beq
(3.29\pm 0.33)\times 10^{-4}~~~({\rm SM~Theory}).
\eeq
This $10\%$ error, whether one interprets it as a $1\sigma$ error
or $95\%$ C.L., clearly implies that there should be comparable error
in the theory prediction of any theory evaluated at the same NLO
rigor.  Supersymmetry, it should be noted, has not been calculated
fully to NLO.  Therefore it is safe to presume that the supersymmetry
prediction will be at least 10\% uncertain, and it must be taken
into account in any careful analysis.

Equally important as the accuracy of the theory prediction
is the fact that the experimental
measurement of $B(b\to s\gamma)$ is not a pure observable in that
a severe cut on the photon energy is needed to reduce charm backgrounds
in the analysis at CLEO.  This introduces 
theoretical uncertainties~\cite{Falk:1998cs,Kagan:1999ym} 
in addition to the obvious ones, such
as imprecise knowledge of the $b$-quark mass, $\alpha_s$ and the
not-completely-known
contributions scaling as $m_c^2/\Lambda_{\rm QCD}^2$.
Therefore, the CLEO measurement is expressed as~\cite{Ahmed:1999fh}
\beq
(3.15\pm 0.35 \pm 0.32 \pm 0.26) \times 10^{-4}~~~({\rm CLEO})
\eeq
where the errors are statistical, systematic, and model-dependence,
respectively. These kinds of varied errors should give pause when
advocating a hard cut on $B(b\to s\gamma)$ in supersymmetry, 
and one should be wary about
deleting any part of parameter space based on 
an apparent incompatibility with the $B(b\to s\gamma)$ constraint.

In our more general MSSM discussion of section 2, the $B(b\to s\gamma)$
constraint does
not even need to be discussed since no values of the squarks masses enter.  
For this more general model, we can simply claim that the
squark masses are sufficiently massive as to contribute little to $B(b\to
s\gamma)$. From the discussions above, we conclude that $B(b\to s\gamma)$,
as with all other observables, usually adds no significant burden in a
quest to find a supersymmetric explanation for $\asusy$.

\bigskip \noindent {\it Acknowledgements:}   SPM is supported in part by
the National Science Foundation grant number PHY-9970691, and JDW
is supported in part by the Department of Energy and the Alfred P. Sloan
Foundation.




\begin{thebibliography}{20}

\bibitem{Czarnecki:2001pv}
For an historical survey of the calculation, see
A.~Czarnecki and W.~J.~Marciano,
``The Muon Anomalous Magnetic Moment: A Harbinger For New Physics,''
hep-ph/0102122.

\bibitem{Brown:2001mg}
H.~N.~Brown {\it et al.}  [Muon g-2 Collaboration],
``Precise Measurement of the Positive Muon Anomalous Magnetic Moment,''
hep-ex/0102017.

\bibitem{Haber:1985rc}
H.~E.~Haber and G.~L.~Kane,
``The Search For Supersymmetry: Probing Physics Beyond The Standard Model,''
Phys.\ Rept.\ {\bf 117}, 75 (1985).

\bibitem{Martin:1997ns}
S.~P.~Martin,
``A supersymmetry primer,''
hep-ph/9709356.

\bibitem{fayet} P.~Fayet, ``Supersymmetry, particle physics and
gravitation" in {\it Unification of the Fundamental Particle
Interactions}, eds. S.~Ferrara, J.~Ellis, and P. van Nieuwenhuizen
(Plenum, New York, 1980), p. 587.

\bibitem{Grifols:1982vx}
J.~A.~Grifols and A.~Mendez,
``Constraints On Supersymmetric Particle Masses From (G-2) Mu,''
Phys.\ Rev.\ D {\bf 26}, 1809 (1982).

\bibitem{Ellis:1982by}
J.~Ellis, J.~Hagelin and D.~V.~Nanopoulos,
``Spin 0 Leptons And The Anomalous Magnetic Moment Of The Muon,''
Phys.\ Lett.\ B {\bf 116}, 283 (1982).

\bibitem{Barbieri:1982aj}
R.~Barbieri and L.~Maiani,
``The Muon Anomalous Magnetic Moment In Broken Supersymmetric Theories,''
Phys.\ Lett.\ B {\bf 117}, 203 (1982).

\bibitem{Kosower:1983yw}
D.~A.~Kosower, L.~M.~Krauss and N.~Sakai,
``Low-Energy Supergravity And The Anomalous Magnetic Moment Of The Muon,''
Phys.\ Lett.\ B {\bf 133}, 305 (1983).

\bibitem{Yuan:1984ww}
T.~C.~Yuan, R.~Arnowitt, A.~H.~Chamseddine and P.~Nath,
``Supersymmetric Electroweak Effects On G-2 (Mu),''
Z.\ Phys.\ C {\bf 26}, 407 (1984).

\bibitem{Vendramin:1989rd}
I.~Vendramin,
``Constraints On Supersymmetric Parameters From Muon Magnetic Moment,''
Nuovo Cim.\ A {\bf 101}, 731 (1989).

\bibitem{Abel:1991dv}
S.~A.~Abel, W.~N.~Cottingham and I.~B.~Whittingham,
``The Muon magnetic moment in flipped SU(5),''
Phys.\ Lett.\ B {\bf 259}, 307 (1991).

\bibitem{Lopez:1994vi}
J.~L.~Lopez, D.~V.~Nanopoulos and X.~Wang,
``Large (g-2)-mu in $SU(5) x U(1)$ supergravity models,''
Phys.\ Rev.\ D {\bf 49}, 366 (1994)
[hep-ph/9308336].

\bibitem{Moroi:1996yh}
T.~Moroi,
``The Muon Anomalous Magnetic Dipole Moment in 
the Minimal Supersymmetric Standard Model,''
Phys.\ Rev.\ D {\bf 53}, 6565 (1996)
[hep-ph/9512396].

\bibitem{Carena:1997qa}
M.~Carena, G.~F.~Giudice and C.~E.~Wagner,
``Constraints on supersymmetric models from the muon anomalous 
magnetic  moment,''
Phys.\ Lett.\ B {\bf 390}, 234 (1997)
[hep-ph/9610233].

\bibitem{Ibrahim:2000hh}
T.~Ibrahim and P.~Nath,
``CP violation and the muon anomaly in N = 1 supergravity,''
Phys.\ Rev.\ D {\bf 61}, 095008 (2000)
[hep-ph/9907555].

\bibitem{Ibrahim:2000aj}
T.~Ibrahim and P.~Nath,
``Effects of large CP violating phases on g(mu)-2 in MSSM,''
Phys.\ Rev.\ D {\bf 62}, 015004 (2000)
[hep-ph/9908443].

\bibitem{Cho:2000sf}
G.~Cho, K.~Hagiwara and M.~Hayakawa,
``Muon g-2 and precision electroweak physics in the MSSM,''
Phys.\ Lett.\ B {\bf 478}, 231 (2000)
[hep-ph/0001229].

\bibitem{Chattopadhyay:2000ws}
U.~Chattopadhyay, D.~K.~Ghosh and S.~Roy,
``Constraining anomaly mediated supersymmetry breaking 
framework via on  going muon g-2 experiment at Brookhaven,''
Phys.\ Rev.\ D {\bf 62}, 115001 (2000)
[hep-ph/0006049].

\bibitem{Brignole:1999gf}
We do not discuss here superlight gravitino 
contributions to $g-2$.  For more information on this interesting
possibility, see A.~Brignole, E.~Perazzi and F.~Zwirner,
``On the muon anomalous magnetic moment in models 
with a superlight  gravitino,''
JHEP{\bf 9909}, 002 (1999)
[hep-ph/9904367].

\bibitem{Anderson:2000ui}
G.~Anderson, H.~Baer, C.~Chen and X.~Tata,
``The reach of Fermilab Tevatron upgrades for 
SU(5) supergravity models  with non-universal gaugino masses,''
Phys.\ Rev.\ D {\bf 61}, 095005 (2000)
[hep-ph/9903370].

\bibitem{Abel:2000vs}
S.~Abel {\it et al.}  [SUGRA Working Group Collaboration],
``Report of the SUGRA working group for run II of the Tevatron,''
hep-ph/0003154.

\bibitem{Giudice:1999bp}
G.~F.~Giudice and R.~Rattazzi,
``Theories with gauge-mediated supersymmetry breaking,''
Phys.\ Rept.\ {\bf 322}, 419 (1999)
[hep-ph/9801271].

\bibitem{Culbertson:2000am}
R.~Culbertson {\it et al.},
``Low-scale and gauge-mediated supersymmetry breaking at the 
Fermilab  Tevatron Run II,''
hep-ph/0008070.

\bibitem{Martin:1997zb}
S.~P.~Martin,
``Generalized messengers of supersymmetry breaking and the 
sparticle mass  spectrum,''
Phys.\ Rev.\ D {\bf 55}, 3177 (1997)
[hep-ph/9608224].


\bibitem{Degrassi:1998es}
G.~Degrassi and G.~F.~Giudice,
``QED logarithms in the electroweak corrections to the muon 
anomalous  magnetic moment,''
Phys.\ Rev.\ D {\bf 58}, 053007 (1998)
[hep-ph/9803384].


\bibitem{Drees:1986vd}
M.~Drees,
``Intermediate Scale Symmetry Breaking And The Spectrum Of 
Super Partners In Superstring Inspired Supergravity Models,''
Phys.\ Lett.\ B {\bf 181}, 279 (1986).

\bibitem{Hagelin:1990ta}
J.~S.~Hagelin and S.~Kelley,
``Sparticle Masses As A Probe Of GUT Physics,''
Nucl.\ Phys.\ B {\bf 342}, 95 (1990).
A.~E.~Faraggi, J.~S.~Hagelin, S.~Kelley and D.~V.~Nanopoulos,
``Sparticle spectroscopy,''
Phys.\ Rev.\ D {\bf 45}, 3272 (1992).

\bibitem{Lleyda:1993xf}
A.~Lleyda and C.~Munoz,
``Nonuniversal soft scalar masses in supersymmetric theories,''
Phys.\ Lett.\ B {\bf 317}, 82 (1993)
[hep-ph/9308208].

\bibitem{Kawamura:1994yv}
Y.~Kawamura and M.~Tanaka,
``Scalar mass spectrum as a probe of E(6) gauge symmetry breaking,''
Prog.\ Theor.\ Phys.\ {\bf 91} (1994) 949.
Y.~Kawamura, H.~Murayama and M.~Yamaguchi,
``Probing symmetry breaking pattern using sfermion masses,''
Phys.\ Lett.\ B {\bf 324}, 52 (1994)
[hep-ph/9402254].

\bibitem{Kolda:1996iw}
C.~Kolda and S.~P.~Martin,
``Low-energy supersymmetry with D term contributions to scalar masses,''
Phys.\ Rev.\ D {\bf 53}, 3871 (1996)
[hep-ph/9503445].

\bibitem{Wells:1994cu}
J.~D.~Wells, C.~Kolda and G.~L.~Kane,
``Implications of $\Gamma(Z \to b \bar b)$ 
for supersymmetry searches and model building,''
Phys.\ Lett.\ B {\bf 338}, 219 (1994)
[hep-ph/9408228].

\bibitem{Baer:1999sp}
H.~Baer, F.~E.~Paige, S.~D.~Protopopescu and X.~Tata,
``ISAJET 7.48: A Monte Carlo event generator for p p, anti-p p, and  
$e^+ e^-$ reactions,''
hep-ph/0001086.

\bibitem{Kane:1994td}
G.~L.~Kane, C.~Kolda, L.~Roszkowski and J.~D.~Wells,
``Study of constrained minimal supersymmetry,''
Phys.\ Rev.\ D {\bf 49}, 6173 (1994)
[hep-ph/9312272].

\bibitem{Abreu:2001fw}
See, for example, P.~Abreu {\it et al.}  [DELPHI Collaboration],
``Search for the standard model Higgs boson at LEP in the year 2000,''
Phys.\ Lett.\ B {\bf 499}, 23 (2001)
[hep-ex/0102036].

\bibitem{Hall:1994gn}
L.~J.~Hall, R.~Rattazzi and U.~Sarid,
``The Top quark mass in supersymmetric SO(10) unification,''
Phys.\ Rev.\ D {\bf 50}, 7048 (1994)
[hep-ph/9306309].

\bibitem{DeRujula:1990fe}
A.~De Rujula, S.~L.~Glashow and U.~Sarid,
``Charged Dark Matter,''
Nucl.\ Phys.\ B {\bf 333}, 173 (1990).

\bibitem{Everett:2001tq}
L.~Everett, G.~L.~Kane, S.~Rigolin and L.~Wang,
``Implications of Muon g-2 for Supersymmetry and 
for Discovering Superpartners Directly,''
hep-ph/0102145.

\bibitem{Feng:2001tr}
J.~L.~Feng and K.~T.~Matchev,
``Supersymmetry and the Anomalous Anomalous 
Magnetic Moment of the Muon,''
hep-ph/0102146.

\bibitem{Komine:2001fz}
S.~Komine, T.~Moroi and M.~Yamaguchi,
``Recent Result from E821 Experiment on Muon g-2 and 
Unconstrained Minimal Supersymmetric Standard Model,''
hep-ph/0102204.

\bibitem{Baer:1994ew}
H.~Baer, C.~Chen, F.~Paige and X.~Tata,
``Detecting sleptons at hadron colliders and supercolliders,''
Phys.\ Rev.\ D {\bf 49}, 3283 (1994)
[hep-ph/9311248].

\bibitem{Kane:2000kc}
G.~L.~Kane, S.~F.~King and L.~Wang,
``What will we learn if a Higgs boson is found?,''
hep-ph/0010312.

\bibitem{Ellis:2000sv}
J.~Ellis, G.~Ganis, D.~V.~Nanopoulos and K.~A.~Olive,
``What if the Higgs boson weighs 115-GeV?,''
hep-ph/0009355.

\bibitem{Ellis:1991nz}
J.~Ellis, G.~Ridolfi and F.~Zwirner,
``Radiative corrections to the masses of supersymmetric Higgs bosons,''
Phys.\ Lett.\ B {\bf 257}, 83 (1991).

\bibitem{Haber:1991aw}
H.~E.~Haber and R.~Hempfling,
``Can the mass of the lightest Higgs boson of the minimal 
supersymmetric model be larger than m(Z)?,''
Phys.\ Rev.\ Lett.\ {\bf 66}, 1815 (1991).

\bibitem{Drees:2000bs}
M.~Drees, Y.~G.~Kim, T.~Kobayashi and M.~M.~Nojiri,
``Direct detection of neutralino dark matter and the 
anomalous dipole  moment of the muon,''
hep-ph/0011359.

\bibitem{Baltz:2001ts}
E.~A.~Baltz and P.~Gondolo,
``Implications of muon anomalous magnetic moment for 
supersymmetric dark matter,''
hep-ph/0102147.

\bibitem{Chattopadhyay:2001vx}
U.~Chattopadhyay and P.~Nath,
``Upper limits on sparticle masses from g-2 and the possibility 
for  discovery of SUSY at colliders and in dark matter searches,''
hep-ph/0102157.

\bibitem{Arnowitt}
R. Arnowitt, B. Dutta, B. Hu, Y. Santoso,
``Muon g-2, Dark Matter Detection and Accelerator Physics,''
hep-ph/0102344.

\bibitem{Bernabei:2000qi}
R.~Bernabei {\it et al.}  [DAMA Collaboration],
``Search for WIMP annual modulation signature: 
Results from DAMA / NaI-3 and DAMA / NaI-4 
and the global combined analysis,''
Phys.\ Lett.\ B {\bf 480}, 23 (2000).

\bibitem{Accomando:2000eg}
For example, see
E.~Accomando, R.~Arnowitt, B.~Dutta and Y.~Santoso,
``Neutralino proton cross sections in supergravity models,''
Nucl.\ Phys.\ B {\bf 585}, 124 (2000)
[hep-ph/0001019].

\bibitem{Wells:1998ci}
J.~D.~Wells,
``Supersymmetric dark matter with a cosmological constant,''
Phys.\ Lett.\ B {\bf 443}, 196 (1998)
[hep-ph/9809504].

\bibitem{Ellis:2001ms}
J.~Ellis, T.~Falk, G.~Ganis, K.~A.~Olive and M.~Srednicki,
``The CMSSM parameter space at large tan beta,''
hep-ph/0102098.

\bibitem{Ellis:2000mm}
J.~Ellis, T.~Falk, K.~A.~Olive and M.~Srednicki,
``Calculations of neutralino stau coannihilation channels and the  
cosmologically relevant region of MSSM parameter space,''
Astropart.\ Phys.\ {\bf 13}, 181 (2000)
[hep-ph/9905481].

\bibitem{Gherghetta:1999sw}
T.~Gherghetta, G.~F.~Giudice and J.~D.~Wells,
``Phenomenological consequences 
of supersymmetry with anomaly-induced  masses,''
Nucl.\ Phys.\ B {\bf 559}, 27 (1999)
[hep-ph/9904378].

\bibitem{Moroi:2000zb}
T.~Moroi and L.~Randall,
``Wino cold dark matter from anomaly-mediated SUSY breaking,''
Nucl.\ Phys.\ B {\bf 570}, 455 (2000)
[hep-ph/9906527].

\bibitem{Jeannerot:1999yn}
R.~Jeannerot, X.~Zhang and R.~Brandenberger,
``Non-thermal production of neutralino cold dark matter from 
cosmic  string decays,''
JHEP{\bf 9912}, 003 (1999)
[hep-ph/9901357].

\bibitem{Baer:1998jq}
H.~Baer, M.~Brhlik, D.~Castano and X.~Tata,
``$b\to s \gamma$ constraints on the minimal supergravity model 
with large  tan(beta),''
Phys.\ Rev.\ D {\bf 58}, 015007 (1998)
[hep-ph/9712305].

\bibitem{Carena:2001uj}
M.~Carena, D.~Garcia, U.~Nierste and C.~E.~Wagner,
``$b\to s \gamma$ and supersymmetry with large tan(beta),''
Phys.\ Lett.\ B {\bf 499}, 141 (2001)
[hep-ph/0010003].

\bibitem{Degrassi:2000qf}
G.~Degrassi, P.~Gambino and G.~F.~Giudice,
``$B \to X/s \gamma$ in supersymmetry: Large contributions 
beyond the  leading order,''
JHEP{\bf 0012}, 009 (2000)
[hep-ph/0009337].

\bibitem{Ellis:2001yu}
J.~Ellis, D.~V.~Nanopoulos and K.~A.~Olive,
``Combining the Muon Anomalous Magnetic Moment with other 
Constraints on the CMSSM,''
hep-ph/0102331.

\bibitem{Dimopoulos:1997yq}
S.~Dimopoulos, S.~Thomas and J.~D.~Wells,
``Sparticle spectroscopy and electroweak symmetry breaking with  
gauge-mediated supersymmetry breaking,''
Nucl.\ Phys.\ B {\bf 488}, 39 (1997)
[hep-ph/9609434].

\bibitem{Randall:1999uk}
L.~Randall and R.~Sundrum,
``Out of this world supersymmetry breaking,''
Nucl.\ Phys.\ B {\bf 557}, 79 (1999)
[hep-th/9810155].

\bibitem{Giudice:1998xp}
G.~F.~Giudice, M.~A.~Luty, H.~Murayama and R.~Rattazzi,
``Gaugino mass without singlets,''
JHEP{\bf 9812}, 027 (1998)
[hep-ph/9810442].

\bibitem{Choi} K.~Choi, K.~Hwang, S.K.~Kang, K.Y.~Lee, W.Y.~Song,
``Probing the messenger of supersymmetry breaking 
by the muon anomalous magnetic moment", [hep-ph/0103048].

\bibitem{Ahmed:1999fh}
S.~Ahmed {\it et al.}  [CLEO Collaboration],
``$b \rightarrow s$ gamma branching fraction and CP asymmetry,''
hep-ex/9908022.

\bibitem{Falk:1998cs}
A.~F.~Falk,
``Heavy quark effective theory and inclusive B decays,''
Nucl.\ Instrum.\ Meth.\ A {\bf 408}, 7 (1998)
[hep-ph/9712364].

\bibitem{Kagan:1999ym}
A.~L.~Kagan and M.~Neubert,
``{QCD} anatomy of $B \rightarrow  X/s$ gamma decays,''
Eur.\ Phys.\ J.\ C {\bf 7}, 5 (1999)
[hep-ph/9805303].


\end{thebibliography}
\end{document}